# Representation learning for time-domain high-energy astrophysics: Discovery of extragalactic fast X-ray transient XRT 200515

Steven Dillmann [1,2]★† Juan Rafael Martínez-Galarza [3] Roberto Soria [4,5] Rosanne Di Stefano [3] and Vinay L. Kashyap [3]

[1] *Institute for Computational and Mathematical Engineering, Stanford University, Stanford, CA 94305, USA*
[2] *Department of Physics, Cavendish Laboratory, University of Cambridge, Cambridge CB3 0HE, UK*
[3] *Center for Astrophysics | Harvard & Smithsonian, Cambridge, MA 02138, USA*
[4] *INAF – Osservatorio Astrofisico di Torino, Strada Osservatorio 20, I-10025 Pino Torinese, Italy*
[5] *Sydney Institute for Astronomy, School of Physics A28, The University of Sydney, Sydney, NSW 2006, Australia*

## ABSTRACT

We present a novel representation learning method for downstream tasks like anomaly detection, unsupervised classification, and similarity searches in high-energy data sets. This enabled the discovery of a new extragalactic fast X-ray transient (FXT) in *Chandra* archival data, XRT 200515, a needle-in-the-haystack event and the first *Chandra* FXT of its kind. Recent serendipitous discoveries in X-ray astronomy, including FXTs from binary neutron star mergers and an extragalactic planetary transit candidate, highlight the need for systematic transient searches in X-ray archives. We introduce new event file representations, $E - t$ maps and $E - t - dt$ cubes, that effectively encode both temporal and spectral information, enabling the seamless application of machine learning to variable-length event file time series. Our unsupervised learning approach employs PCA or sparse autoencoders to extract low-dimensional, informative features from these data representations, followed by clustering in the embedding space with DBSCAN. New transients are identified within transient-dominant clusters or through nearest-neighbour searches around known transients, producing a catalogue of 3559 candidates (3447 flares and 112 dips). XRT 200515 exhibits unique temporal and spectral variability, including an intense, hard <10 s initial burst, followed by spectral softening in an ∼800 s oscillating tail. We interpret XRT 200515 as either the first giant magnetar flare observed at low X-ray energies or the first extragalactic Type I X-ray burst from a faint, previously unknown low-mass X-ray binary in the LMC. Our method extends to data sets from other observatories such as *XMM–Newton*, *Swift-XRT*, *eROSITA*, *Einstein Probe*, and upcoming missions like *AXIS*.

**Key words:** methods: data analysis – software: machine learning – stars: magnetars – stars: peculiar – X-rays: bursts – (transients:) gamma-ray bursts.

## 1 INTRODUCTION

Recent serendipitous discoveries, such as extragalactic fast X-ray transients (FXTs) linked to neutron star merger candidates as electromagnetic counterparts to gravitational wave events (Lin et al. 2022) and an X-ray dip associated with the first extragalactic planet candidate (Di Stefano et al. 2021), underscore the challenges of identifying such rare events within large X-ray catalogues. Beyond magnetar-powered FXTs as the aftermath of binary neutron star mergers (Dai et al. 2006; Metzger, Quataert & Thompson 2008; Zhang 2013; Bauer et al. 2017; Sun, Zhang & Gao 2017; Xue et al. 2019), other interesting origins of extragalactic FXTs include supernova shock breakouts (Soderberg et al. 2008; Modjaz et al. 2009; Alp & Larsson 2020; Novara et al. 2020), tidal disruption events (Jonker et al. 2013) including quasi-periodic eruptions (Arcodia et al. 2021; Chakraborty et al. 2021), or binary self-lensing events (D'Orazio & Di Stefano 2018, 2020; Hu et al. 2020). Both because of their very stochastic nature and because of narrow-field X-ray missions such as the Chandra X-ray Observatory (*Chandra*) (Weisskopf et al. 2000), *XMM–Newton* (Jansen et al. 2001), and *Swift-XRT* (Burrows et al. 2005) are not designed as wide time-domain surveys, X-ray transient discoveries are often serendipitous. They can be found in observations that were originally proposed for a completely unrelated science objective, and are rarely the target of the observation. In many cases, serendipitously discovered X-ray sources do not get characterized or classified, since their transient nature is not immediately obvious. Instead, observations with X-ray transients often get stored in large data archives and remain unnoticed. This raises the need for a systematic search for short-duration phenomena in high-energy catalogues. New missions such as *eROSITA* (Predehl et al. 2021), *Einstein Probe* (Yuan et al. 2022), and the upcoming *AXIS* Observatory (Reynolds et al. 2024) target X-ray transients more directly, thus the development of novel transient detection methods is becoming even more relevant. The temporary, unpredictable and 'unusual' nature of X-ray transients distinguishes them from 'normal' X-ray source emissions. From a data science perspective, they can be understood as 'anomalies' within a large data set. Existing methods for identifying X-ray transients primarily rely on statistical tests of variability (Yang et al. 2019; Pastor-Marazuela et al. 2020; Quirola-Vásquez et al. 2022; Quirola-Vásquez et al. 2023). While effective within specific constraints, these approaches are inherently limited by their underlying assumptions, which may not capture the diverse nature of transient phenomena. In contrast, machine learning offers a more flexible, expressive, and scalable framework, making it particularly well-suited for anomaly detection in large, high-dimensional data sets with diverse transient types. While optical time-domain surveys are at the forefront of leveraging extensive observational programmes, like *ZTF* (Bellm et al. 2019) or the upcoming *LSST* survey (Ivezić et al. 2019), and neural network-based anomaly detection tools to identify rare sources among countless ordinary objects (Martínez-Galarza et al. 2021; Villar et al. 2021; Muthukrishna et al. 2022), the X-ray astronomy community has only recently begun exploring the potential of machine learning to classify sources (Yang et al. 2022; Pérez-Díaz et al. 2024) or to search for needle-in-a-haystack events in large X-ray data sets and archives (Kovačević et al. 2022; Dillmann & Martínez-Galarza 2023). The effectiveness of machine learning methods largely depends on the algorithm's ability to learn useful representations from the data.

*Representation learning* (Bengio, Courville & Vincent 2013) is an increasingly popular technique in astronomy used in supervised, semi-supervised, self-supervised, and unsupervised frameworks (Naul et al. 2018; Hayat et al. 2021; Walmsley et al. 2022; Mohale & Lochner 2024; Slijepcevic et al. 2024). It involves creating or learning meaningful representations for specific modalities of scientific data, which can then be used for downstream tasks such as in this work, anomaly detection and similarity searches. The compressed representations live in a low-dimensional embedding space, in which anomalous data samples are well-separated from more ordinary ones.

We propose a new unsupervised representation learning method to perform a large-scale search for X-ray transients in the *Chandra* archive. High-energy catalogues include individual X-ray source observations in the form of event files. The variable length of these time series poses a challenge in creating consistent representations suitable for transient searches with machine learning. Most deep learning algorithms take a fixed-length input for all data samples. In order to effectively represent event files over a broad range of lengths, we introduce novel fixed-length event file representations, which take into account both their time-domain and energy-domain information. Applying feature extraction and dimensionality reduction techniques, for example with sparse autoencoders, we create a representation space that encodes scientifically meaningful information, such as the spectral and variability properties of the astrophysical sources. Previously identified X-ray transients occupy distinct, well-isolated clusters in the embedding space. Using clustering techniques and nearest-neighbour searches allows us to effectively explore these transient-dominant clusters to discover new X-ray transients. We collect the identified X-ray flare and dip candidates in a publicly available catalogue, serving as a fertile ground for new discoveries in time-domain high-energy astrophysics.

Among these candidates, we identify an intriguing extragalactic FXT, XRT 200515, which exhibits unique temporal and spectral characteristics distinct from any previously reported *Chandra* FXTs. The transient's initial hard < 10 s burst shows a sharp rise exceeding four orders of magnitude, followed by spectral softening in an ∼800 s oscillating tail. This transient is likely related to either a giant magnetar flare (GMF) from a distant soft gamma repeater (SGR) behind the Large Magellanic Cloud (LMC) or an extragalactic Type

I X-ray burst from a faint low-mass X-ray binary (LMXB) in the LMC. Each of these interpretations presents its own set of challenges. Alternatively, XRT 200515 could be a new type of astronomical phenomenon found by our machine learning approach.

Our method is the first unsupervised representation learning approach for transient searches in high-energy astrophysics. It is applicable to data sets from high-energy catalogues like *Chandra*, *XMM–Newton*, *Swift-XRT*, *eROSITA*, and *Einstein Probe*. We created semantically meaningful representations that were also used for regression and classification (Dillmann & Martínez-Galarza 2023). These can later be aligned with other data modalities, such as optical images or infrared spectra to design multimodal models (Mishra-Sharma, Song & Thaler 2024; Parker et al. 2024; Rizhko & Bloom 2024; The Multimodal Universe Collaboration 2024; Zhang et al. 2024) using contrastive learning (Radford et al. 2021), that can improve on current state-of-the-art methods used to characterize the associated objects. Ultimately, this work and other representation and contrastive learning approaches lay the groundwork for developing large-scale foundation models in astronomy.

The paper is organized as follows. In Section 2, we provide information on the data set of *Chandra* event files used in this analysis. In Section 3, we describe in detail the implementation of our novel transient detection approach leveraging representation learning. In Section 4, we present and discuss the results in form of the semantically meaningful representation space of the event files, the catalogue of X-ray transient candidates and the discovery of the new *Chandra* transient XRT 200515. Finally, we highlight our contributions to time-domain high-energy astrophysics and outline potential directions for extending this work in the future in Section 5.

The relevant code, a demonstration of the pipeline, and an interactive embedding selection, transient search and light curve plotting tool are available online at the GitHub repository https://github.com/StevenDillmann/ml-xraytransients-mnras.

## 2 DATA SET

We use data from the Chandra Source Catalog (CSC) version 2.1 (Evans et al. 2024), which includes all publicly available X-ray sources detected by *Chandra* as of December 2021. For this study, we focus specifically on observations from the Advanced CCD Imaging Spectrometer (ACIS). CSC 2.1 had not been fully released at the time our analysis was performed, but catalogue data was available for sources that had completed processing in the *Current Database View*,[1] a snapshot of which we took on 2023 April 11. CSC 2.1 performs source detection on stacked observations, and catalogue properties are provided both for these stack-level detections, and for each of observation-level detection that contribute to a stack detection. Because we are interested in short-time variability that happens within a *single* observation of a source, we use the catalogue products for the observation-level detections in our analysis. For a given X-ray detection, two types of products are provided in the CSC: (i) data base tables with source properties, such as fluxes in the different X-ray energy bands, hardness ratios, variability indices, etc., and (ii) file-based data products for each detection of a source, such as the detect regions, the *Chandra* point spread function (PSF) at that location, etc. The following observation-level catalogue properties are relevant for our analysis:

(i) var_prob_b: The probability that a source detection is variable in time for the broad energy band (0.5–7 keV), as estimated

---
[1] https://cxc.cfa.harvard.edu/csc2/

using the Gregory–Loredo algorithm (Gregory & Loredo 1992). In this paper, we call this quantity $p_{var}^b$.

(ii) var_index_b: The variability index in the broad-band, which indicates the level of confidence for time variability. A variability index of 6 or larger indicates variability at a confidence of at least $2\sigma$. In this paper, we call this quantity $I_{var}^b$.

(iii) hard_<hs/ms/hm>: The hardness ratios, which quantify the relative fraction of photons detected in two given bands chosen between the soft (0.5–1.2 keV), medium (1.2–2 keV), and hard (2–7 keV) bands for a source detection. For example, a value of the hard-to-soft hardness ratio close to 1 indicates that most of the photons detected are in the hard energy band, whereas a value close to −1 indicates that most photons are detected in the soft band. In this paper, we call these quantities $HR_{hs}$, $HR_{ms}$, and $HR_{hm}$.

From the catalogue data products available for observation-level X-ray detections, we are interested in the region event file. This event file consists of a list of all individual photon events detected in a small bounding box around a source detection, listing their energies, arrival times, and detector coordinates. These multivariate time series are the basis for the characterization of an X-ray source: light curves, spectra, images, coordinates, and other properties are derived from the distribution of the listed quantities. In this analysis, we directly use these event files as our primary data products. The values of the catalogue properties listed above serve as summary statistics for the detection associated with a given region event file. We only include event files with more than five events and a signal-to-noise ratio above five to minimize spurious signals from low-number statistics in faint sources. We also exclude detections that are flagged for pile-up,[2] i.e. those with a pileup fraction larger than 5 per cent, which corresponds to a maximum pileup warning of 0.1 in CSC 2.1. For the resulting detections, we filter the event files to include only events contained within the detection region for each source. These detection regions are also provided as data products in CSC 2.1, and consist of the ellipse that includes the 90 per cent encircled counts fraction of the PSF at the source location. Due to the low background level in *Chandra* observations, the majority of events selected after this spatial filtering are expected to be events associated with the X-ray source, not the background. In the selected event files, we only include photon events within good time intervals (GTIs), which are time periods of valid, high-quality data. No other pre-processing is required. The final data set consists of 95 473 filtered event files from 58 932 sources, resulting in an average of 1.62 observations per source. This includes 9003 new sources that have been added as part of the CSC 2.1 release, in addition to the sources from the previous release.

## 3 METHODS

In this work, we introduce a novel representation learning based anomaly detection method to systematically search for X-ray transients in high-energy archives. We begin with an overview of the method here and provide detailed explanations of each step in individual subsections. The full pipeline is illustrated in Fig. 1. Starting with the event files described in Section 2, we (i) build two novel and uniform event file representations by binning their arrival times and energies into $E - t$ maps (event file representation I) or $E - t - dt$ cubes (event file representation II); (ii) use principal component analysis (PCA; feature extraction I) or sparse autoencoders (feature extraction II) to extract informative features from the event file representations; (iii) apply dimensionality reduction to the extracted features to create a low-dimensional embedding space; (iv) use density-based clustering to create embedding clusters that group event files with similar characteristics, for example transient behaviour or certain spectral features. Previously identified transients like the extragalactic magnetar-powered flare candidate reported by Lin et al. (2022) and the extragalactic planet candidate dip reported by Di Stefano et al. (2021), shown in Fig. 2, occupy well-isolated clusters in the embedding space. Exploring these clusters and conducting nearest-neighbour searches enables us to effectively find analogues to bona-fide time-domain anomalies, while at the same time grouping them according to their spectral properties. We compile the identified transient candidates in a catalogue. While our approach is designed and tested using *Chandra* data, it is applicable to any data set consisting of event lists, like those from other high-energy telescopes. The described transient detection approach is applied to both types of event file representations with both feature extraction methods, resulting in four different embeddings. We denote the different cases as described in Table 1.

### 3.1 Event file representation

The different event files in the data set are variable in length $N$ and duration $T$, as shown in Appendix A. The large variation in the number of events and duration highlights the challenge in producing uniform data representations that preserve relevant information on time variability and spectral properties. While there exist machine learning architectures that take variable length inputs, the significant differences in the number of events from object to object make standardization of the inputs challenging, even when these architectures are used (Martínez-Galarza & Makinen 2022). As a first step in our analysis, we introduce two-dimensional and three-dimensional fixed-length representations based on an informed binning strategy for the event files, similar to the DMDT maps for optical light curves introduced by Mahabal et al. (2017).

### 3.1.1 2D histogram representation ($E - t$ maps)

Assume an event file with $N$ photons and a photon arrival time column $t$ with entries $\{t_k\}_{k=1}^N$ and energy column $E$ with entries $\{E_k\}_{k=1}^N$. The event file duration is given by $T = t_N - t_1$. The energy column entries take values in the broad energy band of *Chandra*'s ACIS instrument, i.e. $E_k \in [E_{min}, E_{max}]$, where $E_{min} = 0.5$ keV and $E_{max} = 7$ keV comes from considering appropriate boundaries for the energy response of *Chandra*'s ACIS instrument. Beyond these boundaries, the telescope's aperture effective area is low for the majority of detected sources. First, we obtain the normalized time column, given by $\tau = \frac{t - t_1}{T}$, and the logarithm of the energy column, given by $\epsilon = \log E$. The resulting boundaries for normalized time column are $\tau \in [\tau_{min}, \tau_{max}]$, where $\tau_{min} = 0$ and $\tau_{max} = 1$. The range for the log-energy column is $\epsilon \in [\epsilon_{min}, \epsilon_{max}]$, where $\epsilon_{min} = \log 0.5$ keV and $\epsilon_{max} = \log 7$ keV.

Next, we determine the dimensionality of our representations. For each event file, we determine the optimal number of bins in the energy dimension, $n_\epsilon$, with the Freedman–Diaconis rule (Freedman & Diaconis 1981), a widely used method that balances the trade-off between too noisy histograms (too many bins) and not informative enough histograms (too few bins). The optimal bin width

---

[2]Photon pileup occurs when multiple photons interact with the same detection cell in a single CCD frame. Pileup leads to a decrease in the observed count rate and skews the spectrum towards higher energies (Davis 2001).

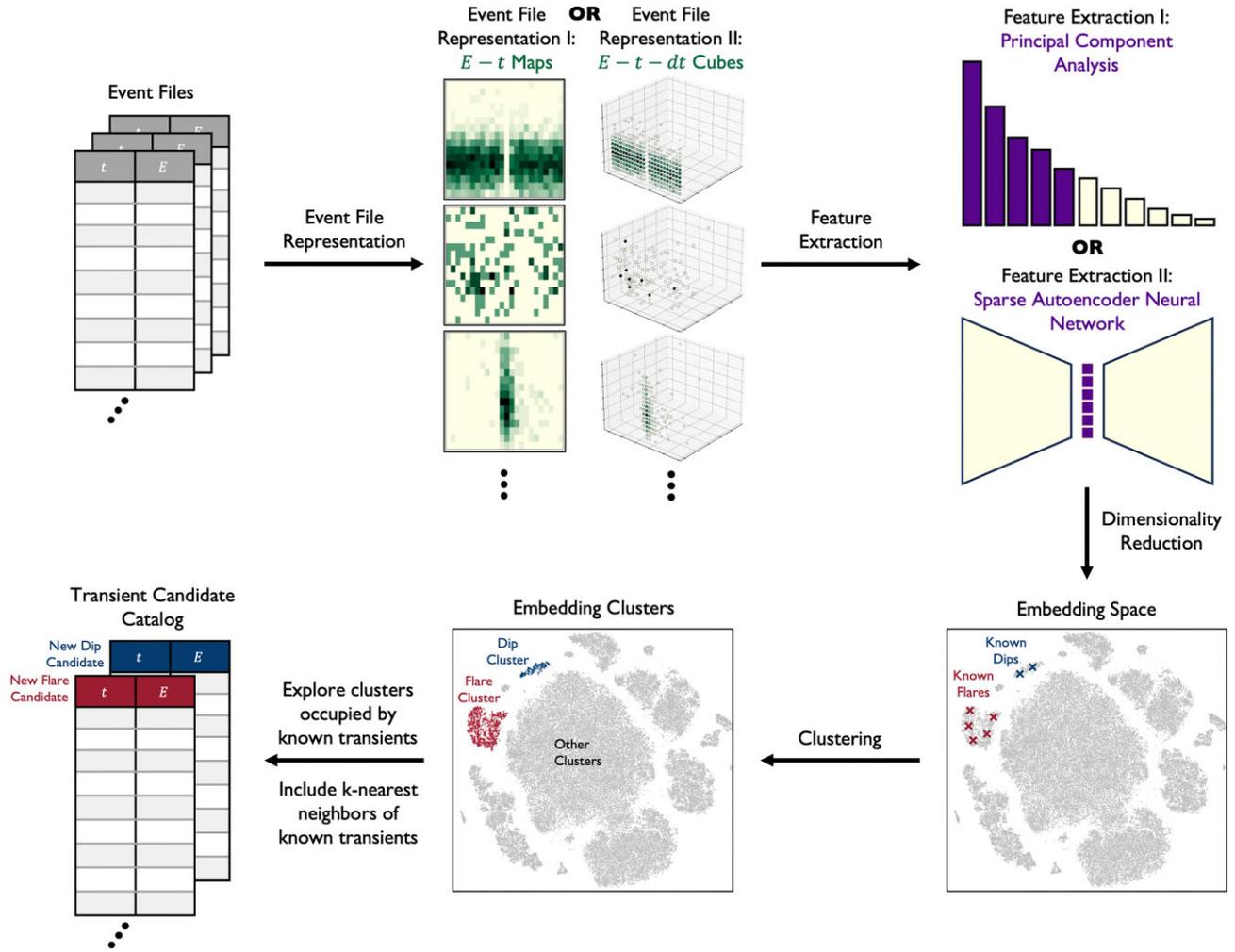

**Figure 1.** Flowchart of the proposed representation learning approach for anomaly detection and similarity searches in time-domain high-energy astrophysics, enabling the systematic detection of transients in high-energy archives. The first step is to create uniform event file representations by binning photon arrival times and energies in the event files into $E - t$ maps (event file representation I) or $E - t - dt$ cubes (event file representation II). The second step involves extracting informative features from these event representations via PCA (feature extraction I) or sparse autoencoders (feature extraction II). The third step is to apply dimensionality reduction to the extracted features and to create a low-dimensional embedding space, which is clustered in the fourth step using density-based clustering. Previously identified transients occupy well-isolated clusters on the edges of the embedding space, thus new transients can be identified by exploring these transient-dominant edge clusters and performing nearest-neighbour searches around known bona-fide flares and dips. Finally, we compile these search results in a publicly available catalogue of transient candidates, serving as a fertile ground for the discovery of new X-ray transients.

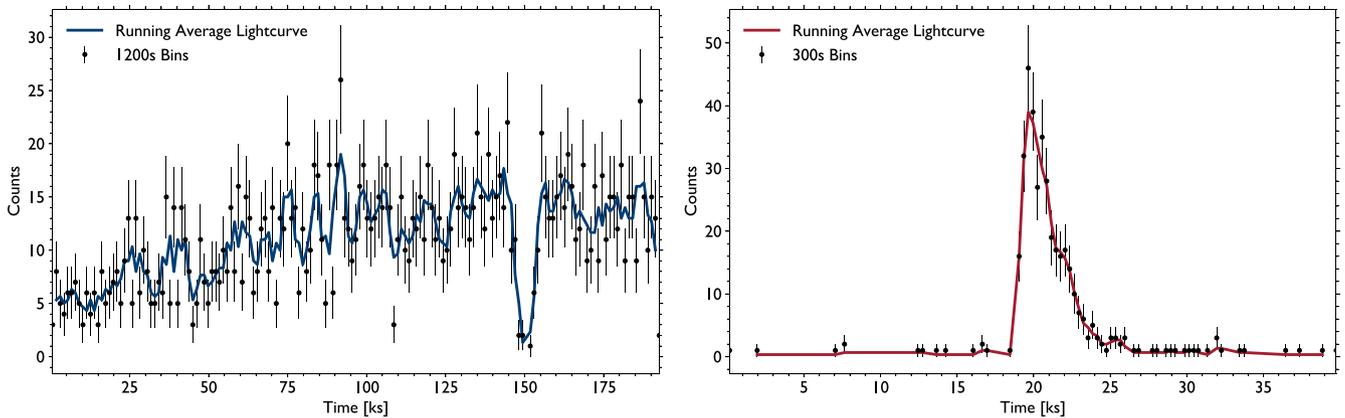

**Figure 2.** *Left panel:* Light curve of the first extragalactic planet candidate dip reported by Di Stefano et al. (2021) detected in the *Chandra* observation ObsID 13814. *Right panel:* Light curve of the magnetar-powered X-ray flare candidate reported by Lin et al. (2022) detected in the *Chandra* observation ObsID 4062.

**Table 1.** Naming of the different embedding result cases based on the event file representation and feature extraction method.

| Case | Event file representation | Feature extraction |
|---|---|---|
| *2D-PCA* | $E - t$ maps | PCA |
| *3D-PCA* | $E - t - dt$ cubes | PCA |
| *2D-AE* | $E - t$ maps | Sparse autoencoder |
| *3D-AE* | $E - t - dt$ cubes | Sparse autoencoder |

$b_\epsilon$ according to this rule is calculated in the following way:

$$b_\epsilon = 2 \frac{\text{IQR}(\epsilon)}{N^{\frac{1}{3}}}, \tag{1}$$

where $\text{IQR}(\epsilon)$ represents the inter-quartile range of the $\epsilon$ values for a given event file of length $N$. Subsequently, we obtain the optimal number of energy bins $n_\epsilon$ with

$$n_\epsilon = \frac{\epsilon_{\max} - \epsilon_{\min}}{b_\epsilon}. \tag{2}$$

For each event file, we determine the optimal number of bins in the time dimension, $n_\tau$, with the help of the Bayesian Blocks algorithm, which was specifically developed for time series analysis in astronomy (Scargle et al. 2013). This algorithm partitions the time series into adaptive width bins or blocks that are statistically distinct from neighbouring blocks; that is, within a given time-ordered Bayesian block, events grouped in that block are consistent with having a similar event arrival rate. We use the default ASTROPY implementation of Bayesian blocks, and set the false alarm probability parameter to $p_0 = 0.01$ (Astropy Collaboration 2013), which implies a 1 per cent probability of declaring a change of rate when there is none. For each event file, we define the optimal uniform bin width $b_\tau$ as the minimum bin width calculated by the Bayesian Blocks algorithm, and then find the optimal number of time bins $n_\tau$ with

$$n_\tau = \frac{\tau_{\max} - \tau_{\min}}{b_\tau}. \tag{3}$$

The optimal number of bins is different for each event file, due to their different lengths $N$ and durations $T$. To select a bin size that can be applied to all event files, we consider the distributions of these optimal bin sizes, which are shown in Fig. 3. For the distribution of $n_\tau$ values we only use those event files for which $p_{\text{var}}^b > 0.9$. The intent of this is to effectively capture variability time-scales that are associated with short time-domain events, such as flares and dips.

We choose the 90th percentile value of each distribution to set the final number of bins in each dimension. That is, only 10 per cent of the event files will have an optimal number of bins that is larger than the chosen values $n_\epsilon = 16$ and $n_\tau = 24$. The choice of the 90th percentile, rather than the mean or mode, is motivated by the need to capture sufficient statistical detail even for long event files, while keeping the size of the resulting representations computationally tractable. Choosing a lower resolution would risk losing significant details in the representation, particularly short-duration events such as flares and dips within longer event files. The $E - t$ maps are the 2D histogram representations with size $(n_\tau, n_\epsilon) = (24, 16)$ that result from binning the events according to the optimized number of bins.

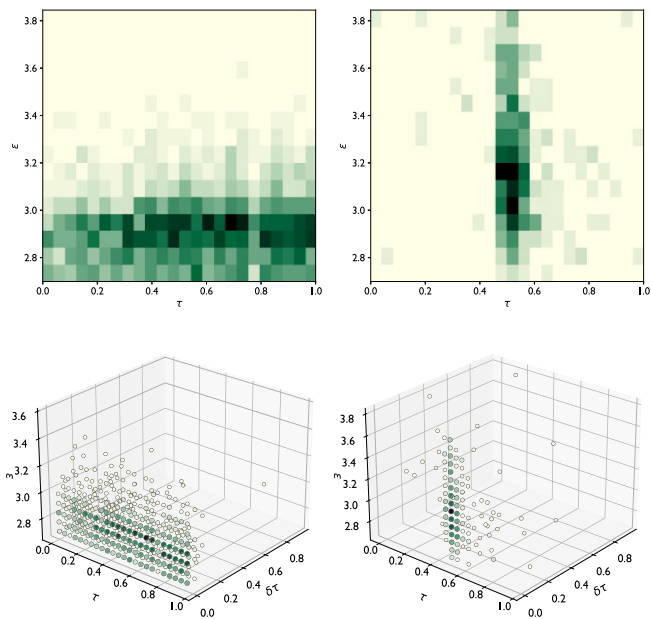

**Figure 4.** *Upper panel:* 2D histogram representations ($E - t$ maps) for the extragalactic dip in Di Stefano et al. (2021) (left) and extragalactic flare in Lin et al. (2022) (right). *Lower panel:* 3D histogram representations ($E - t - dt$ cubes) for the same events. Darker bins indicate higher counts, while lighter bins indicate lower counts.

Fig. 4 shows the $E - t$ maps for the known extragalactic dip reported by Di Stefano et al. (2021) and known extragalactic flare reported by Lin et al. (2022).

### 3.1.2 3D histogram representation ($E - t - dt$ cubes)

We now introduce the $E - t - dt$ cubes, which extend the $E - t$ maps by a third dimension that serves as a proxy for the photon arrival rate. For an event file of length $N$, consider the array of time differences between consecutive photon arrivals $\boldsymbol{\Delta t}$ with entries $\Delta t_k = t_{k+1} - t_k$ for $k = 1, 2, \ldots, N - 1$. We again scale and normalize the obtained values, so that they adopt values between 0 and 1, using in each case the minimum value $\Delta t_{\min}$ and maximum

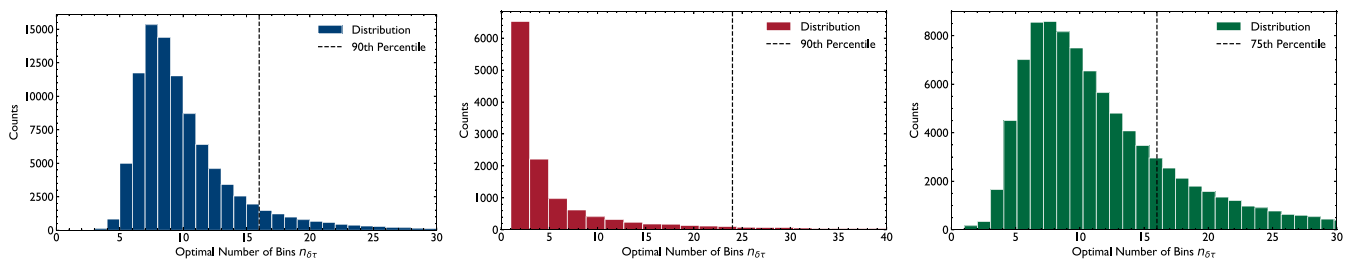

**Figure 3.** The distribution of the optimal number of bins for the energy dimension $n_\epsilon$ (left), time dimension $n_\tau$ (middle), time difference dimension $n_{\delta\tau}$ (right). The distribution of $n_\tau$ only includes event files for which $p_{\text{var}}^b > 0.9$. The vertical lines indicate the number of bins chosen for the event file representations.

value $\Delta t_{max}$. This provides the third dimension $\delta\tau$:

$$\delta\tau = \frac{\Delta t - \Delta t_{min}}{\Delta t_{max} - \Delta t_{min}}. \tag{4}$$

The additional dimension is intended to better isolate short-duration features in time variability by capturing high photon arrival rates, which are typical of flares, as well as very low photon arrival rates, which are typical of dips. The boundaries of our histogram representations in this dimension are $\delta\tau \in [\delta\tau_{min}, \delta\tau_{max}]$, where $\delta\tau_{min} = 0$ and $\delta\tau_{max} = 1$. We determine the optimal number of bins in the $\delta\tau$ dimension, $n_{\delta\tau}$, again by computing the optimal bin width $b_{\delta\tau}$ with the Freedman–Diaconis rule and dividing the range for $\delta\tau$ by $b_{\delta\tau}$:

$$b_{\delta\tau} = 2\frac{\mathrm{IQR}(\delta\tau)}{N^{\frac{1}{3}}}, \tag{5}$$

$$n_{\delta\tau} = \frac{\delta\tau_{max} - \delta\tau_{min}}{b_{\delta\tau}}. \tag{6}$$

The distribution of $n_{\delta\tau}$ across the event files is shown in Fig. 3. Most of the relevant time-domain information is already captured by $\tau$, but adding $\delta\tau$ provides an additional marker for dips and flares that can be shorter than the time-scales probed by our chosen binning of $\tau$.

Unlike in the other two dimensions, we choose the 75th percentile value of the distribution as our final choice of common binning, which results in $n_{\delta\tau} = 16$. This is because in order to identify short transients, we need to capture strong deviations in $\delta\tau$ only. Choosing a lower value for $n_{\delta\tau}$ reduces noise an improves computational tractability. Having both $\tau$ and $\delta\tau$ represented also breaks any assumption of stationarity, in that we can be sensitive to transient events happening at any time during the observation of the source, and break degeneracies between periodic and non-periodic features in the representations presented by Martínez-Galarza & Makinen (2022). The $E - t - dt$ cubes are the resulting 3D histogram event file representations with size $(n_\tau, n_\epsilon, n_{\delta\tau}) = (24, 16, 16)$.

Fig. 4 shows the $E - t - dt$ cubes for the known extragalactic dip reported by Di Stefano et al. (2021) and known extragalactic flare reported by Lin et al. (2022).

### 3.1.3 Feature notation

The event file representations can now be used as inputs for various statistical learning and machine learning algorithms. For the $i$th event file in the data set of length $m = 95\,473$, we denote the corresponding feature vector as $x_i = [x_1, x_2, \ldots, x_n]_i$, where $n = n_\tau \cdot n_\epsilon = 384$ for the $E - t$ maps and $n = n_\tau \cdot n_\epsilon \cdot n_{\delta\tau} = 6144$ for the $E - t - dt$ cubes. The set of all feature vectors is denoted as $\mathsf{X} = [x_1, x_2, \ldots, x_m]^\top$ with size $(m, n)$.

## 3.2 Feature extraction I: principal component analysis

We use PCA (Pearson 1901) provided by SCIKIT-LEARN (Pedregosa et al. 2011), as our first feature extraction method. The extracted principal components should encode relevant time-domain and spectral information of the event file they represent. PCA involves transforming a data set into a new coordinate system by finding the principal components of the data that capture most of the variance in the data. By projecting the data set on to principal components, PCA reduces the dimensionality of the data while retaining the most important information, which increases the interpretability of high-dimensional data.

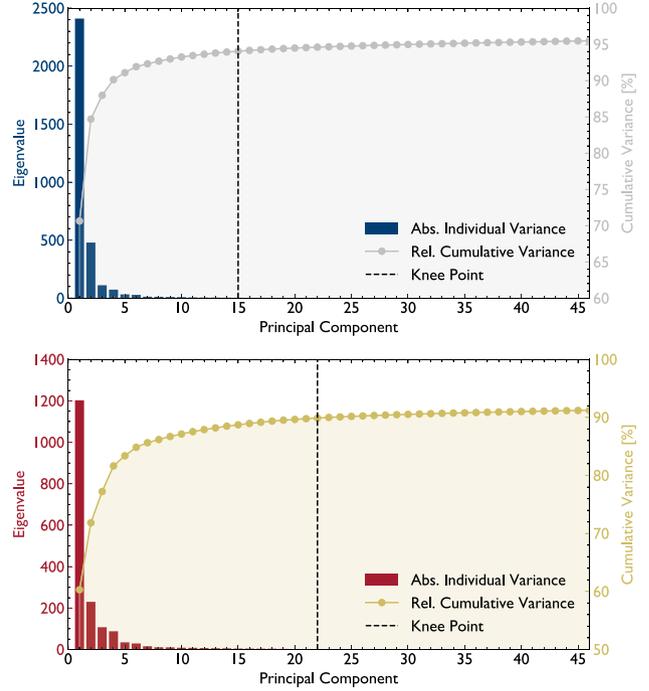

**Figure 5.** Scree plot for the principal components of the $E - t$ maps (top) and $E - t - dt$ cubes (bottom). The scree plots show the amount of variance explained by each individual principal component including the knee point in the cumulative variance.

### 3.2.1 PCA algorithm

We start with the feature vector set $\mathsf{X}$ of size $(m, n)$ representing our data set with $m$ samples and $n$ dimensions. PCA aims to find a new coordinate system defined by a set of orthogonal axes, i.e. the principal components, that captures the maximum amount of variance in the data. The PCA result is a transformed data set $\mathsf{X}_{pc}$ obtained by projecting $\mathsf{X}$ on to the principal components:

$$\mathsf{X}_{pc} = \mathsf{X}\mathsf{W}, \tag{7}$$

where $\mathsf{W}$ is matrix of size $(n, n_{pc})$ containing the first $n_{pc}$ principal components to be retained as its columns and $\mathsf{X}_{pc}$ is of size $(m, n_{pc})$ with a reduced dimensionality of $n_{pc}$. For a more detailed explanation of the algorithm, we refer the reader to Jolliffe (2002).

### 3.2.2 Principal components retained

The main PCA hyperparameter is the number of principal components $n_{pc}$ to retain. Fig. 5 shows two scree plots illustrating the amount of variance explained by each principal component in descending order and the cumulative proportion of variance explained by the principal components for both $E - t$ maps and $E - t - dt$ cubes. A common approach to determine the optimal value of $n_{pc}$ is to find the knee point in the cumulative scree plot of the principal components. This balances the objective of minimizing the dimensionality while retaining as much information as possible. Defining the knee point as the point beyond which adding additional principal components increases the amount of variance by less than 0.1 per cent gives $n_{pc} = 15$ for $E - t$ maps and $n_{pc} = 22$ for $E - t - dt$ cubes as indicated in Fig. 5. These capture 94.1 per cent and 89.9 per cent of the variance, respectively.

**Table 2.** Summary of the encoder architecture of the convolutional autoencoder used to extract informative features from the $E - t$ maps. Note that each layer has a leaky ReLU activation function and that each standard convolutional layer is followed by batch normalization with momentum 0.9.

| Layer | Output shape | Filters | Kernel | Stride |
|---|---|---|---|---|
| Input | (24, 16) | – | – | – |
| Convolution | (24, 16) | 32 | (3, 3) | – |
| Convolution | (12, 8) | 32 | (2, 2) | 2 |
| Convolution | (12, 8) | 16 | (3, 3) | – |
| Convolution | (6, 4) | 16 | (2, 2) | 2 |
| Flatten | 384 | – | – | – |
| Dense | 192 | – | – | – |
| Dense | 48 | – | – | – |
| Dense (bottleneck) | 12 | – | – | – |

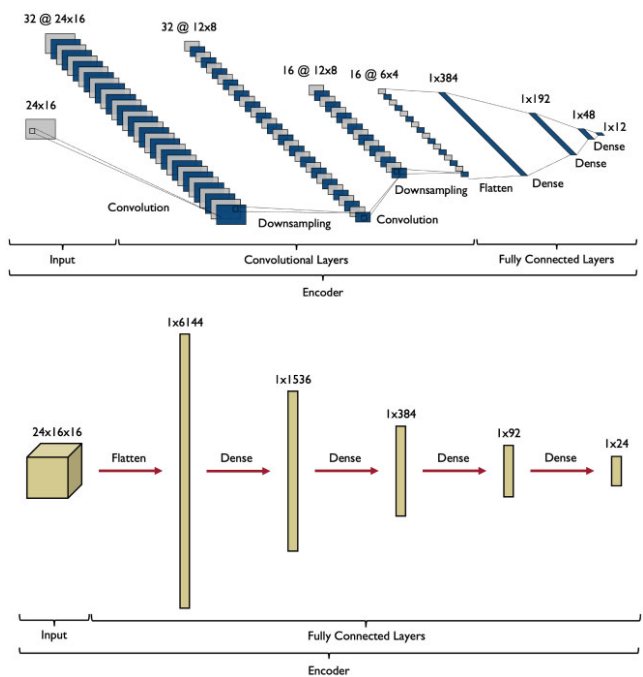

**Figure 6.** Encoder architecture of the convolutional autoencoder used for the $E - t$ maps (top) and of the fully connected autoencoder used for the $E - t - $ d$t$ cubes (bottom). The decoder architecture is simply a mirror image of the encoder.

### 3.3 Feature extraction II: sparse autoencoder neural network

As an alternative to PCA, we now build autoencoder (Hinton & Salakhutdinov 2006) models with TENSORFLOW (Abadi et al. 2016) to learn a set of latent features from the $E - t$ maps and $E - t - $ d$t$ cubes that can be used to isolate transients and encode specific spectral properties. An autoencoder is composed of two neural networks, an encoder and a decoder, which work together to learn a compressed representation of the input data. The encoder network takes the input data and maps it to a lower-dimensional representation, often called 'latent space' or 'bottleneck'. The number of neurons in the bottleneck determines the dimensionality of the learned representation. The decoder network then aims to reconstruct the original input from this compressed representation. The decoder is typically a mirrored version of the encoder gradually upsampling the latent space until the output matches the dimensions of the original input. By minimizing the reconstruction error between input and output during training, the model learns a low-dimensional representation of the input. The bottleneck forces the encoder to capture the most important information necessary for accurate reconstruction, effectively compressing the input and learning to extract informative features in an unsupervised manner. Once the autoencoder is trained, the encoder network can be used as a stand-alone feature extractor to obtain a compressed representation of the input data, which can be used for downstream tasks such as clustering or anomaly detection. As opposed to PCA, which is a linear technique that works well for linearly correlated data but fails to capture complex non-linear relationships, an autoencoder is able to learn complex non-linear relationships. We design two different autoencoders to process the $E - t$ maps and $E - t - $ d$t$ cubes.

#### 3.3.1 Convolutional autoencoder

In a convolutional autoencoder (Masci et al. 2011), both the encoder and decoder network consist of convolutional layers (LeCun et al. 1998), which perform convolutions over the input using a filter. These filters are small matrix kernels with learnable weights that slide across the input, allowing the network to capture high-level features while preserving important spatial hierarchies and relationships, which is why they are often used for image-like data. This makes this architecture particularly well-suited to recognize spatial patterns such as dips or flares in our $E - t$ maps. To gradually reduce the dimension of the input while it is being passed through the encoder network, we use stride convolution layers (Simonyan & Zisserman 2014) with a stride value of 2 for downsampling. This means that the learnable filter jumps two pixels at a time as it slides over the input. The output of the convolutional layers is a feature map, which is then flattened to a feature vector and passed through a series of fully connected layers, where every neuron in the previous layer is connected to every neuron in the next layer. These fully connected layers are responsible for mapping the learned features to a lower dimensional latent representation in the bottleneck and perform non-linear transformations while downsampling through the use of non-linear activation functions. The final latent space has $n_{ae} = 12$ elements, representing the most essential features of the input data, which can now be used for further downstream tasks. Fig. 6 shows a diagram of the encoder part of the model and Table 2 summarizes its architecture.

#### 3.3.2 Fully connected autoencoder

Our $E - t - $ d$t$ cubes introduce an additional dimension resulting in sparse 3D input data. Convolutional layers assume regular grid-like data, making them less effective for handling sparse data. Moreover, very expensive 3D convolutional operations would substantially increase complexity of the model. Therefore, we use a simple fully connected autoencoder for the $E - t - $ d$t$ cubes. Its encoder network consists of a series of fully connected layers, which gradually map the original input data to a latent space with $n_{ae} = 24$ elements. Fig. 6 shows a diagram of the encoder part of the model and Table 3 summarizes its architecture.

#### 3.3.3 Activation functions

Neural networks are able to learn and represent complex non-linear relationships due to the introduction of non-linear activation functions within their layers. An activation function is a mathematical function used in a neural network to determine whether a neuron should be activated or not, based on its input. It essentially decides how much of the input signal should pass through the neuron, producing an output that can either be passed to the next layer or used to make predictions. The popular rectified linear unit (ReLU) activation function ReLU$(x) = \max(0, x)$ (Nair & Hinton 2010) is simple and computationally efficient. To mitigate any potential encounters of the 'dying ReLU problem', where neurons become non-responsive during training, we choose an extended version called Leaky ReLU (Maas et al. 2013):

$$\text{Leaky ReLU}(x) = \max(\alpha x, x), \tag{8}$$

where $\alpha = 0.1$ is a hyperparameter that defines the slope of the function for negative input values. ReLU sets all negative values in the input to zero, while Leaky ReLU allows a small negative slope for negative inputs, which can help prevent neurons from dying. As for the output layer, we want any values to be mapped to a range between 0 and 1, which is achieved by using the sigmoid activation function:

$$\text{sigmoid}(x) = \frac{1}{1 + e^{-x}}. \tag{9}$$

**Table 3.** Summary of the encoder architecture of the fully connected autoencoder used to extract informative features from the $E - t - dt$ cubes. Note that each layer has a Leaky ReLU activation function and that each standard fully connected layer is followed by batch normalization with momentum 0.9.

| Layer | Output shape |
|---|---|
| Input | (24, 16, 16) |
| Flatten | 6144 |
| Dense | 1536 |
| Dense | 384 |
| Dense | 92 |
| Dense (bottleneck) | 24 |

### 3.3.4 Loss function and sparsity regularization

In order to encourage the autoencoder to generate reconstructions close to the original inputs, we use the mean squared error (MSE) as a measure of the reconstruction quality given by

$$\text{MSE} = \frac{1}{m} \sum_{i=1}^{m} (x_i - \hat{x}_i)^2, \tag{10}$$

where $x_i$ is the $i$th element of the input vector and $\hat{x}_i$ is the corresponding is reconstructed output. The MSE is a straightforward measure of reconstruction error, and its differentiability allows efficient gradient computation for updating model weights via gradient-based optimization.

Our neural networks are so-called sparse autoencoders (Ng et al. 2011), which promote sparsity in the learned representation, meaning only a small subset of the neurons in the network are active at any given time. Sparse representations are valuable for our work because they help extract highly informative features from the input, while disregarding irrelevant or noisy information. To encourage sparsity in the latent space, we introduce a L1 regularization term in the objective, resulting in the following loss function:

$$L = \text{MSE} + \lambda \cdot \sum_{j=1}^{n_w} |w_j| = \frac{1}{m} \sum_{i=1}^{m} (x_i - \hat{x}_i)^2 + \lambda \cdot \sum_{j=1}^{n_w} |w_j|, \tag{11}$$

where $\lambda = 0.1$ is the regularization strength and $w_j$ are the individual bottleneck weight values of which there are $n_w$ in total. L1 regularization pushes small weights to zero and thus helps the model prioritize the most significant features of the input data, leading to a semantically meaningful latent space.

### 3.3.5 Training

Starting with the original data set with a $m = 95\,473$ samples and using a test split of 0.1 gives us a training and validation set of length 85 925 and a test set of length 9548. Further using a validation split of 0.2, gives 68 740 samples for training and 17 185 for validation. We run the training process for a maximum of 200 epochs with a batch size of 1024 samples. The initial learning rate was set to 0.01 along with an on plateau learning rate scheduler, which dynamically reduces the learning rate by a factor of 0.1 if the validation loss plateaus for longer than 10 epochs. Reducing the learning rate when a plateau is detected can help escape local minima in the loss surface and converge to a more optimal solution in the parameter space. This scheduler is used in combination with the Adaptive Moment Estimation (Adam) optimizer (Kingma & Ba 2014), which is a stochastic gradient descent algorithm combining the benefits of both adaptive learning rates (Duchi, Hazan & Singer 2011) and momentum-based optimization techniques (Sutskever et al. 2013). Finally, we use an early stopping callback to monitor the validation loss. It automatically interrupts the training process if the validation loss does not improve for 25 epochs and restores the weights of the model to the best observed weights during training. The training process for both autoencoder models is shown in Appendix B. Once the autoencoder is trained, we can use the encoder to transform the original data set $\mathbf{X}$ to the feature vector space $\mathbf{X}_{\text{ae}}$ of size $(m, n_{\text{ae}})$ with a reduced dimensionality of $n_{\text{ae}}$ features.

## 3.4 Dimensionality reduction

Using t-SNE (Maaten & Hinton 2008), short for t-Distributed Stochastic Neighbour Embedding, we create two-dimensional embeddings of the informative features previously extracted from the event file representations using PCA or sparse autoencoders. The t-SNE algorithm is a method used to map the input data onto a low-dimensional embedding space, and is particularly useful for the visualization of clusters and patterns in high-dimensional data sets. Each high-dimensional sample is transformed into a low-dimensional embedding in such a way that similar object are nearby points, while dissimilar objects are distant points in the embedding space. Essentially, it aims to capture the local structure of the data by preserving the pairwise similarities between objects while mapping them to a lower-dimensional embedding space.

### 3.4.1 Algorithm

We use our informative features, $\mathbf{X}_{\text{if}} = \mathbf{X}_{\text{pc}}$ or $\mathbf{X}_{\text{if}} = \mathbf{X}_{\text{ae}}$, as input to the t-SNE algorithm to reduce the data to a two-dimensional embedding, denoted as $\mathbf{Z}$. First, t-SNE creates a probability distribution $P$ for pairs of high-dimensional data points in $\mathbf{X}_{\text{if}}$, assigning higher probabilities to similar pairs and lower probabilities to dissimilar ones. This is done by modelling pairwise similarities using a Gaussian kernel with a specific perplexity parameter, which controls the effective number of neighbours considered for each point. Next, t-SNE defines a similar probability distribution $Q$ for the pairwise similarities in the low-dimensional space $\mathbf{Z}$, modelled using a Student's t-distribution. The goal of t-SNE is to minimize the difference between $P$ and $Q$ using gradient descent, with the Kullback–Leibler (KL) divergence (Kullback & Leibler 1951) as the cost function:

$$D_{\text{KL}}(P \mid Q) = \sum_{i \neq j} P_{ij} \log \frac{P_{ij}}{Q_{ij}}, \tag{12}$$

**Table 4.** Chosen t-SNE hyperparameters for different embedding cases.

| Hyperparameter | 2D-PCA | 3D-PCA | 2D-AE | 3D-AE |
|---|---|---|---|---|
| `perplexity` | 30 | 50 | 40 | 60 |
| `learning_rate` | 80 | 120 | 100 | 180 |
| `n_iter` | 4500 | 3500 | 3000 | 2000 |
| `random_state` | 11 | 11 | 2412 | 12 |

where $P_{ij}$ and $Q_{ij}$ represent pairwise similarities in the high- and low-dimensional spaces, respectively. The algorithm iteratively adjusts the low-dimensional embedding $Z$ to minimize the KL divergence, often requiring hundreds to thousands of iterations for convergence. The result of this optimization is a two-dimensional representation $Z$ of size $(m, 2)$, where similar points in the high-dimensional space are clustered closely together.

### 3.4.2 Hyperparameter optimization

The t-SNE algorithm has a number of important hyperparameters to be tuned. The two most important parameters are the `perplexity` and the `learning_rate`. The `perplexity` parameter controls the balance between capturing the local versus global structure in the data, while the `learning_rate` controls the step size at each iteration of the optimization process. The `n_iter` parameter is the number of iterations. To ensure reproducibility, we set a fixed `random_state`. Our t-SNE hyperparameter optimization approach is detailed in Appendix C. A summary of the final t-SNE hyperparameters is provided in Table 4.

### 3.5 Clustering

The next step is the identification of individual clusters in the embedding space using DBSCAN (Hartigan & Wong 1979), short for Density-Based Spatial Clustering of Applications with Noise. Unlike traditional clustering algorithms such as k-means, DBSCAN does not require the number of clusters to be specified, as it identifies dense regions in the data space based on a density criterion.

### 3.5.1 Algorithm

We use our t-SNE embedding space $Z$ as input to the DBSCAN algorithm, which segments the embedding space into multiple clusters. The DBSCAN algorithm has two main hyperparameters. The `eps` parameter defines the radius of the neighbourhood surrounding each point in the data set, while the `minPts` parameter specifies the minimum number of points required within this neighbourhood for a data point to be classified as a core point. A border point is defined as a point that is in the vicinity of at least one core point but has fewer than `minPts` within its neighbourhood. All other points are considered to be noise points. Clusters are then created from the aggregation of core points and their associated border points, with noise points being categorized as outliers. Fig. 7 visualizes the clustering method.

### 3.5.2 Hyperparameter optimization

Our DBSCAN hyperparameter optimization approach is detailed in Appendix C. A summary of the final t-SNE hyperparameters is provided in Table 5.

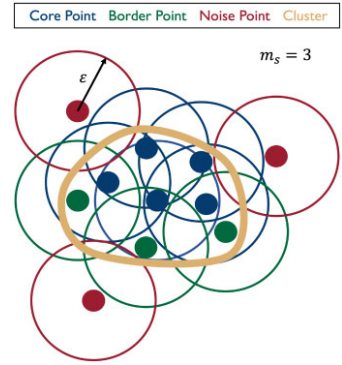

**Figure 7.** Illustration of the DBSCAN clustering algorithm, showing core points as densely connected regions, border points along cluster edges, and noise points as outliers. Adapted from Slipski et al. (2024) with permission.

**Table 5.** Chosen DBSCAN hyperparameters for different embedding cases.

| Hyperparameter | 2D-PCA | 3D-PCA | 2D-AE | 3D-AE |
|---|---|---|---|---|
| `eps` | 2.8 | 2.1 | 1.6 | 1.9 |
| `minPts` | 26 | 23 | 25 | 24 |

### 3.6 Previously reported transients

We highlight the embeddings of previously reported bona-fide transients, listed in Table 6, in our low-dimensional representation space to identify transient-dominant clusters. The flares include extragalactic FXTs reported by Jonker et al. (2013), Glennie et al. (2015), Yang et al. (2019), Lin et al. (2021, 2022), Quirola-Vásquez et al. (2022) and a set of stellar flares found in the data set by manual inspection. The dips include the extragalactic planet candidate in M 51 reported by Di Stefano et al. (2021), the ultraluminous X-ray source (ULX) 2E 1402.4+5440 in NGC 5457 (Colbert & Ptak 2002; Swartz et al. 2004) and the well-studied eclipsing LMXB EXO 0748−676 (Parmar et al. 1986; D'Aì et al. 2014). These transients occupy well-isolated clusters. Exploring transient-dominant clusters and performing nearest-neighbour searches around known transients allows us to find new transients.

### 3.7 Candidate selection

New transients are identified in embedding clusters containing previously reported transients. For well-isolated clusters containing known discovered transients, we use the entire cluster to define new transient candidates. The well-isolated transient-dominant clusters used for candidate selection are listed in Appendix E. However, in a few cases known transients reside within larger poorly separated clusters. Selecting the entire cluster would result in a high number of false positives. To address this, we instead use the *k*-nearest-neighbour (kNN) algorithm (Cover & Hart 1967), identifying the 50 nearest neighbours for each known transient residing in a poorly separated cluster to define additional transient candidates.

### 3.8 Cross-matching

We use an existing cross-match table (Green et al. 2023) between CSC 2.1 and five other catalogues–*Gaia* DR3 (Gaia Collaboration 2021), DESI Legacy Survey DR10 (Dey et al. 2019), PanSTARRS-1 (Chambers et al. 2016), 2MASS (Skrutskie et al. 2006), and the SDSS DR17 catalogue– to complement the X-ray properties with multi-wavelength observations. This includes catalogue identifiers, positions, magnitudes, source type classifications and other columns. We cross-matched our transient candidates with the SIMBAD data base (Wenger et al. 2000) by associating each candidate with the nearest SIMBAD object, provided the object is located within a 5 arcsec radius of the candidate's coordinates listed in the CSC. The multi-wavelength observations of the transient candidates provide valuable information for their characterization and classification.

**Table 6.** Previously reported flares and dips used to identify transient-dominant clusters.

| CSC name | ObsID | Date | Transient type | Description |
|---|---|---|---|---|
| 2CXO J123605.1+622013 | 957 | 2000-02-23 | Flare | Extragalactic fast X-ray transient (Yang et al. 2019) |
| 2CXO J122531.5+130357 | 803 | 2000-05-19 | Flare | Extragalactic fast X-ray transient (Jonker et al. 2013) |
| 2CXO J190725.1+070906 | 1042 | 2001-09-06 | Flare | Unknown origin |
| 2CXO J111128.3+554021 | 2025 | 2001-09-08 | Flare | Extragalactic fast X-ray transient (Quirola-Vásquez et al. 2022) |
| 2CXO J123625.3+621405 | 3389 | 2001-11-21 | Flare | Extragalactic fast X-ray transient (Yang et al. 2019) |
| 2CXO J111908.8−612540 | 2833 | 2002-03-31 | Flare | X-ray source |
| 2CXO J053517.5−051739 | 4395 | 2003-01-08 | Flare | Stellar flare (Orion Variable) |
| 2CXO J053528.1−051856 | 4396 | 2003-01-18 | Flare | Stellar flare (young stellar object) |
| 2CXO J163553.8−472540 | 3877 | 2003-03-24 | Flare | Compact object system (Lin, Webb & Barret 2012) |
| 2CXO J050706.7−315211 | 4062 | 2003-05-10 | Flare | Extragalactic fast X-ray transient (Lin et al. 2022) |
| 2CXO J151457.6+364817 | 3988 | 2003-10-05 | Flare | Stellar flare |
| 2CXO J165334.4−414423 | 6291 | 2005-07-16 | Flare | Magnetic Cataclysmic Variable (Lin, Webb & Barret 2014) |
| 2CXO J025616.7+585756 | 7151 | 2006-06-21 | Flare | Unknown origin |
| 2CXO J074111.5+741450 | 10822 | 2009-06-18 | Flare | Unknown origin |
| 2CXO J112017.5+125818 | 9548 | 2008-03-31 | Flare | Extragalactic fast X-ray transient (Quirola-Vásquez et al. 2022) |
| 2CXO J140828.9−270328 | 12884 | 2011-01-03 | Flare | Extragalactic fast X-ray transient (Glennie et al. 2015) |
| 2CXO J010344.5−214845 | 13454 | 2011-09-19 | Flare | Extragalactic fast X-ray transient (Lin et al. 2022) |
| 2CXO J064114.4+093321 | 13610 | 2011-12-05 | Flare | Stellar flare (RS CVn Variable) |
| 2CXO J064028.7+093059 | 14368 | 2011-12-03 | Flare | Stellar flare (T Tauri Star) |
| 2CXO J064119.6+093144 | 14368 | 2011-12-03 | Flare | Stellar flare (T Tauri Star) |
| 2CXO J095959.4+024646 | 15211 | 2012-12-13 | Flare | Extragalactic fast X-ray transient (Yang et al. 2019) |
| 2CXO J235212.2−464343 | 13506 | 2012-08-30 | Flare | Extragalactic fast X-ray transient (Glennie et al. 2015) |
| 2CXO J030309.0−774435 | 15113 | 2014-03-27 | Flare | Extragalactic fast X-ray transient (Quirola-Vásquez et al. 2022) |
| 2CXO J234503.4−423841 | 20635 | 2017-08-31 | Flare | Extragalactic fast X-ray transient (Lin et al. 2022) |
| 2CXO J134856.4+263944 | 24604 | 2021-04-23 | Flare | Extragalactic fast X-ray transient (Lin et al. 2021) |
| 2CXO J121656.9+374335 | 942 | 2000-05-20 | Dip | Ultraluminous X-ray source |
| 2CXO J140414.2+542604 | 4733 | 2004-05-07 | Dip | Ultraluminous X-ray source (2E 1402.4+5440) |
| 2CXO J140414.2+542604 | 5322 | 2004-05-03 | Dip | Ultraluminous X-ray source (2E 1402.4+5440) |
| 2CXO J140414.2+542604 | 4736 | 2004-11-01 | Dip | Ultraluminous X-ray source (2E 1402.4+5440) |
| 2CXO J140515.6+542458 | 6152 | 2004-11-07 | Dip | Active Galactic Nucleus |
| 2CXO J140414.2+542604 | 6170 | 2004-12-22 | Dip | Ultraluminous X-ray source (2E 1402.4+5440) |
| 2CXO J140414.2+542604 | 4737 | 2005-01-01 | Dip | Ultraluminous X-ray source (2E 1402.4+5440) |
| 2CXO J074833.7−674507 | 9070 | 2008-10-12 | Dip | Low-Mass X-ray binary (EXO 0748−676) |
| 2CXO J074833.7−674507 | 10783 | 2008-10-15 | Dip | Low-Mass X-ray binary (EXO 0748−676) |
| 2CXO J021404.0+275239 | 9550 | 2008-10-03 | Dip | Ultraluminous X-ray source candidate |
| 2CXO J074833.7−674507 | 10871 | 2009-02-25 | Dip | Low-Mass X-ray binary (EXO 0748−676) |
| 2CXO J031702.5−410714 | 11059 | 2010-04-20 | Dip | Low-Mass X-ray binary (EXO 0748−676) |
| 2CXO J031702.5−410714 | 11272 | 2010-05-04 | Dip | Ultraluminous X-ray source (NGC 1291 PSX-2) |
| 2CXO J132939.4+471243 | 13813 | 2012-09-09 | Dip | Hii region |
| 2CXO J132939.9+471236 | 13812 | 2012-09-12 | Dip | Hii region (Di Stefano & Kong 2004) |
| 2CXO J132943.3+471134 | 13814 | 2012-09-20 | Dip | Extragalactic Planet Transit Candidate (Di Stefano et al. 2021) |
| 2CXO J132939.9+471236 | 13814 | 2012-09-20 | Dip | Hii region (Di Stefano & Kong 2004) |

# 4 RESULTS AND DISCUSSION

We now present the results of applying the methods in Section 3 to the set of representations of X-ray event files in the data set from Section 2.

## 4.1 Representation embedding space and clusters

Fig. 8 shows the t-SNE embedding space for the *3D-PCA* and *3D-AE* cases colour-coded by the hardness ratio $HR_{hs}$. The embedding space for the other two cases, *2D-PCA* and *2D-AE*, are shown in Appendix D. The observed hardness ratio gradients in all embedding spaces indicate that the learned representations effectively encode spectral information, in particular at the level of individual clusters, allowing for the identification of X-ray sources with specific spectral signatures. For the *2D-PCA* and *2D-AE* cases, these gradients are more uniform across the embedding space, because the temporal and spectral information of event files are captured by one axis each in the $E - t$ maps. Moreover, some clusters consist exclusively of soft or hard sources, demonstrating that our representations can be leveraged not only to identify transients but also to find analogues to sources with specific spectral characteristics.

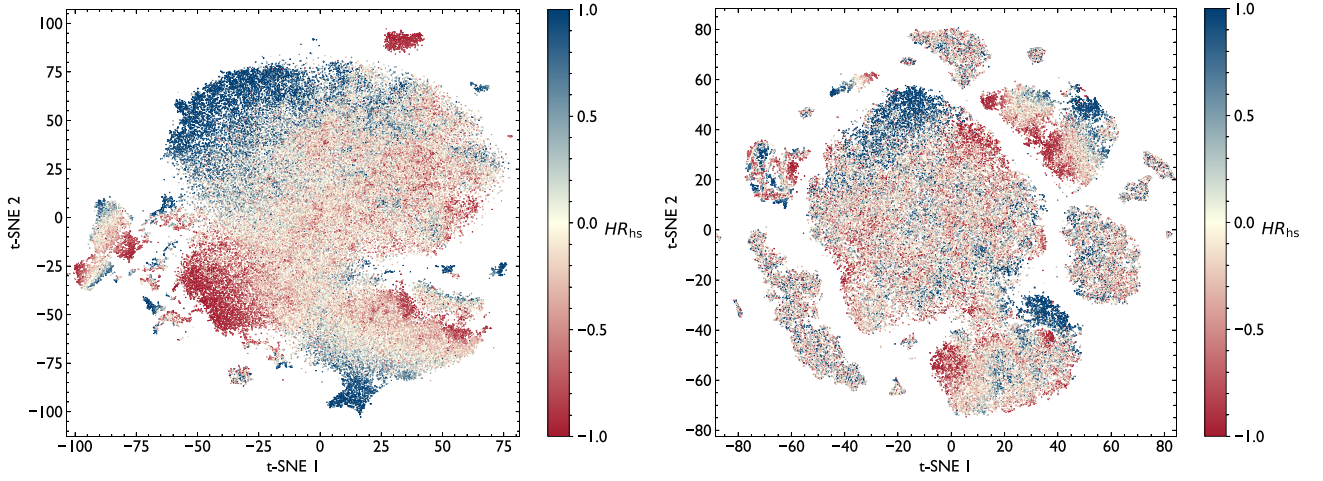

**Figure 8.** Embedding space colour-coded by $HR_{hs}$ for the *3D-PCA* case (left) and *3D-AE* case (right).

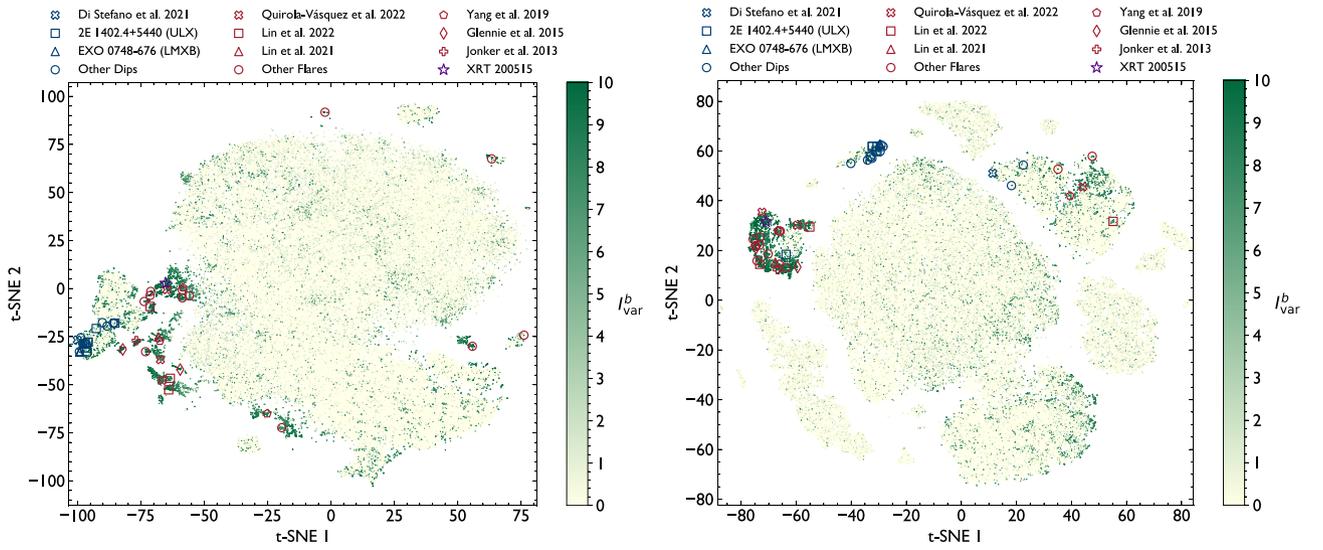

**Figure 9.** Embedding space colour-coded by $I_{var}^{b}$ for *3D-PCA* (left) and *3D-AE* (right). Known transients and XRT 200515 are highlighted.

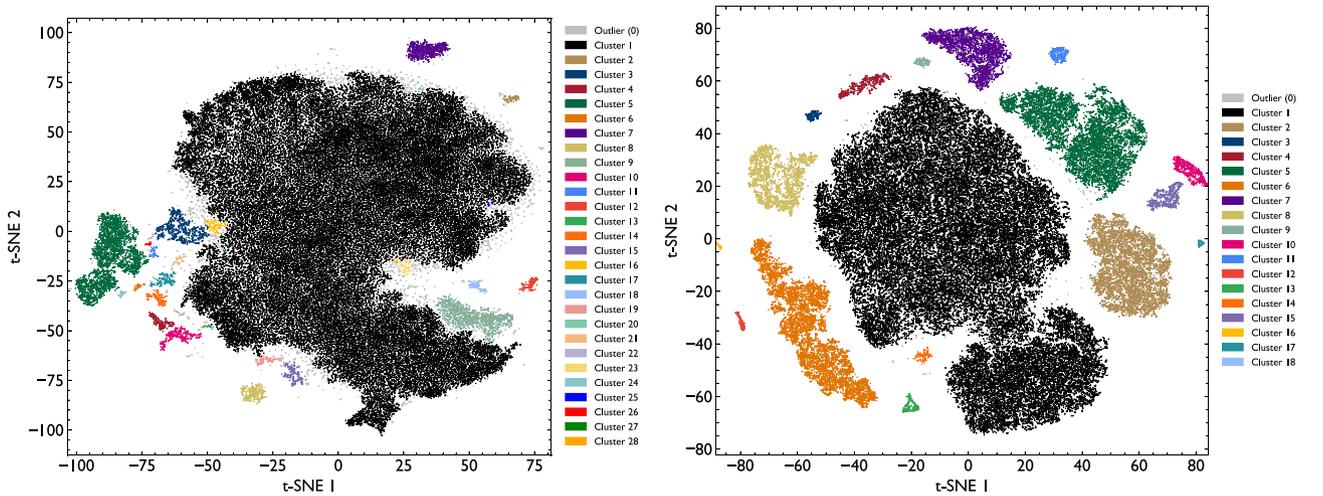

**Figure 10.** Embedding clusters for the *3D-PCA* case (left) and *3D-AE* case (right).

**Table 7.** The first five samples of our transient candidates catalogue showing a subset of selected columns.

| CATALOG_ID | CSC_name | TRANSIENT_TYPE | CSC_ra | CSC_dec | ... | CSC_var_index_b | ... | SIMBAD_otype |
|---|---|---|---|---|---|---|---|---|
| 10049_3 | 2CXO J162636.5−515630 | Dip | 246.652 136 | −51.941 847 | ... | 7.0 | ... | HighMassXBin |
| 10059_961 | 2CXO J174805.3−244656 | Flare | 267.022 413 | −24.782 385 | ... | 8.0 | ... | X |
| 10059_967 | 2CXO J174805.2−244647 | Flare | 267.021 774 | −24.779 915 | ... | 8.0 | ... | LowMassXBin |
| 10062_450 | 2CXO J152010.7−571110 | Flare | 230.044 590 | −57.186 057 | ... | 10.0 | ... | Unknown |
| 10065_31 | 2CXO J170029.9−461310 | Flare | 255.124 760 | −46.219 472 | ... | 9.0 | ... | Unknown |

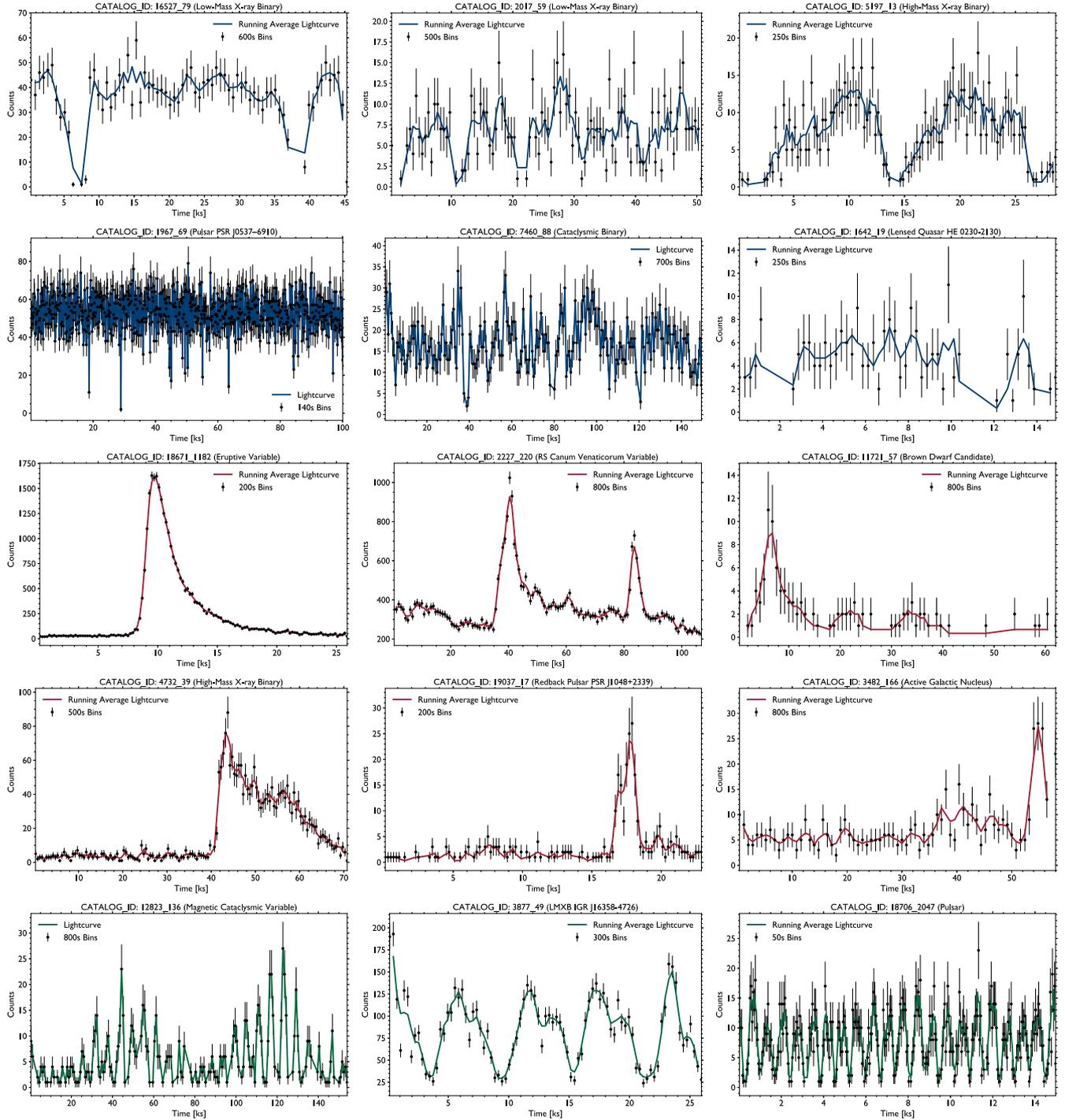

**Figure 11.** Light curves in the 0.5–7 keV energy range for different examples of dips (blue), flares (red) and pulsating or quasi-periodic sources (green) in the transient candidates catalogue. The shown pulsating or quasi-periodic sources are part of the flare candidates.

Fig. 9 shows the *3D-PCA* and *3D-AE* embedding spaces, now colour-coded by the variability index $I_{var}^{b}$ with the other two cases shown in Appendix D. The learned embeddings also encode the temporal behaviour of the sources, with some clusters being dominated by X-ray detections with significant variability, including transient behaviour. To demonstrate this, we also highlight the embeddings of the bona-fide flares and dips listed in Table 6. Note that these occupy very well-defined clusters on the edges of the representation space, allowing for queries of analogue transient behaviour. In the *2D-PCA* and *2D-AE* cases, transient sources are distributed across multiple small clusters on the edges of the embedding spaces. In contrast, the *3D-PCA* and *3D-AE* embedding spaces achieve a significantly more compact clustering of bona-fide transients because temporal features in the event files are given a higher importance by the introduction of an additional time-related axis in the $E - t - dt$ data.

Fig. 10 shows the clusters identified by the DBSCAN algorithm in the *3D-PCA* and *3D-AE* cases. The clusters for the other two cases, *2D-PCA* and *2D-AE*, are shown in Appendix D. The largest cluster in all cases (Cluster 1) corresponds to observations that are not 'anomalous', for example non-variable sources or noisy detections in the low-count regime. We also see multiple smaller clusters on the edges of the embedding space clearly separated from this main cluster. Of special interest are clusters that contain known discovered transients, as these likely host other interesting transients that have not yet been discovered. Some of the edge clusters group observations with similar temporal and spectral behaviour. For example, Cluster 4 in the *3D-PCA* case only contains flares with high hardness ratios. Other clusters instead group observations primarily by similar temporal behaviour, but then show a within-cluster grouping of similar spectral behaviours. For example, Cluster 4 in the *3D-AE* case contains many dipping sources, but show a hardness ratio gradient within the cluster. When comparing the results of different feature extraction methods, we observe that in the *3D-AE* embedding space, nearly all previously identified extragalactic FXTs live within a single, well-isolated cluster (Cluster 8). In contrast, the *3D-PCA* embedding space distributes these extragalactic FXTs across multiple clusters. All of these points underline the effectiveness of our method and that the created representation space is highly informative.

### 4.2 Catalogue of X-ray flare and dip candidates

We identify new transient candidates within clusters that are occupied by previously reported transients and by conducting nearest-neighbour searches around these known transients (Dillmann & Martínez-Galarza 2023). We compile these in a catalogue of X-ray transient candidates, which includes both flares and dips. The selected clusters used to define the new flare and dip candidates in addition to the 50 nearest neighbours of each bona-fide transient are given in Appendix E. Note that from each selected flare cluster, we include only X-ray detections with a variability index $I_{var}^{b} \geq 5$, corresponding to at least 90 per cent confidence in variability per the Gregory–Loredo algorithm, ensuring statistical significance. We also include a select group of interesting sources identified as non-clustered points within the embeddings, particularly pulsating or quasi-periodic sources, to the flare candidates. Lastly, we manually exclude a fraction of false positives identified by visual inspection of the light curves. The resulting catalogue contains a total of 3559 detections (3447 flares and 112 dips), with the catalogue columns described in Appendix F. Table 7 shows the first five samples in our catalogue for a subset of columns. Fig. 11 shows a number of example light curves of the dips and flares in our catalogue. The dip selection shows dips from LMXBs, a high-mass X-ray binary (HMXB), the glitching pulsar PSR J0537−6910 (Marshall et al. 2004), a cataclysmic binary, and the lensed quasar HE 0230−2130 (Wisotzki et al. 1999). The flare selection shows flares from an eruptive variable, a RS CVn variable, a brown dwarf candidate, a HMXB, the redback pulsar PSR J1048+2339 (Deneva et al. 2016) and an active galactic nucleus (AGN). We also show pulsating or quasi-periodic light curves from a magnetic cataclysmic variable, the peculiar LMXB IGR J16358−4726 (Kouveliotou et al. 2003; Patel et al. 2004, 2007) and a pulsar. Fig. 12 shows the distribution of SIMBAD object types in our transient catalogue. About 25 per cent of the transient candidates do not have a SIMBAD match, making them particularly interesting sources for new transient discoveries. Our dip candidates include six *Chandra* observations with prominent dips from the known source CXOGlb J002400.9−720453 in the globular cluster NGC 104 (47 Tuc). The catalogue identifiers for these are `CATALOG_ID: 2737_139, 16527_79, 15747_79, 16529_79, 15748_79, 16528_14`. Our flare candidates include a newly discovered extragalactic FXT, which is characterized and discussed in detail in Section 4.3. Its catalogue identifier is `CATALOG_ID: 23022_122`. We recommend using our catalogue to identify a diverse range of flares and dips. While this work is primarily motivated by the discovery of new extragalactic transients, we intentionally did not exclude galactic stellar flares to enable systematic follow-up studies to study flare incidence rates, the rotational evolution of stars and more. Users interested exclusively in extragalactic transients can filter out galactic sources using metadata from the CSC and the cross-match columns in the catalogue (Dillmann et al. 2025).

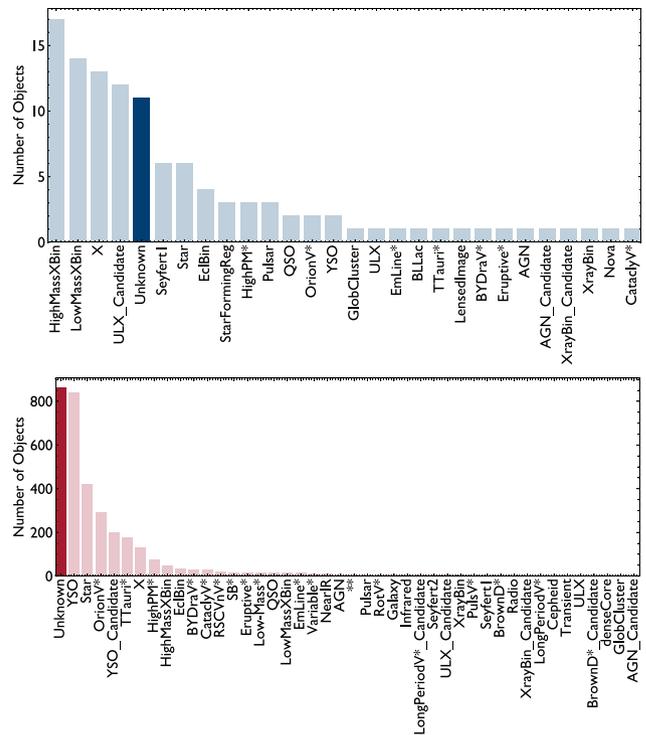

**Figure 12.** Distribution SIMBAD object types in the dip candidates catalogue (blue) and flare candidates catalogue (red). There are 11 dip candidates and 862 flare candidates, for which no SIMBAD match is found.

### 4.3 XRT 200515: A new extragalactic fast X-ray transient

Among the flare candidates in our catalogue, we discovered an intriguing new extragalactic *Chandra* FXT in an observation of the supernova remnant SNR 0509−67.5 in the LMC on 2020 May 15 (Guest et al. 2022). What made this transient stand out from thousands of other flares discovered in this work is the unique temporal variability in its light curve, which exhibits no detectable pre-flare X-ray emission, a sharp rise of at least four orders of magnitude in the count rate to peak intensity followed by a sharp fall, all in a matter of a <10 s, down to ~800 s long oscillating tail. There is also notable spectral variability during the flare, characterized by an initially hard spectrum at the peak, followed by spectral softening in the tail. The combination of these temporal and spectral properties establishes this transient as the first of its kind within the sample of discovered *Chandra* FXTs. We designate this newly discovered FXT as XRT 200515 and present a detailed study and discussion of its potential origins.

#### 4.3.1 X-ray detection by Chandra

The transient XRT 200515 was detected in *Chandra* ObsID 23022. The target of the observation was the supernova remnant SNR 0509−67.5 in the LMC, which is shown in Fig. 13 alongside the newly discovered FXT event. Table 8 summarizes the properties of XRT 200515 and its associated *Chandra* source 2CXO J051117.2−672556 in Obs ID 23022. The transient was captured by the ACIS camera in the S4 chip, and is located significantly off-axis in this observation, at an angular distance of 11.75 arcmin from the aimpoint in the S3 chip. This leads to an elongated and relatively large PSF, which, in this case, is advantageous as it substantially reduces photon pile-up in the initial spike, by spreading the counts over many pixels. We processed the data of *Chandra* observation ObsID 23022 with the Chandra Interactive Analysis of Observations (CIAO) version 4.15 (Fruscione et al. 2006), with calibration data base version 4.9.8. In particular, we created a new level-2 event file with the CIAO task chandra_repro and filter it in energy and time with dmcopy. We obtained the sky position in Table 8 using the CIAO tool wavdetect. To reduce background noise and improve the determination of the source centroid, we applied wavdetect on an image filtered to include only the time interval from the beginning of the flare ($t_0$) until a time $t_0 + 920$ s. The 90 per cent uncertainty radius of 2.0 arcsec is the combination of the uncertainty in the source centroid position reported by wavdetect, and the absolute astrometry uncertainty in a typical ACIS observation for off-axis sources.[3]

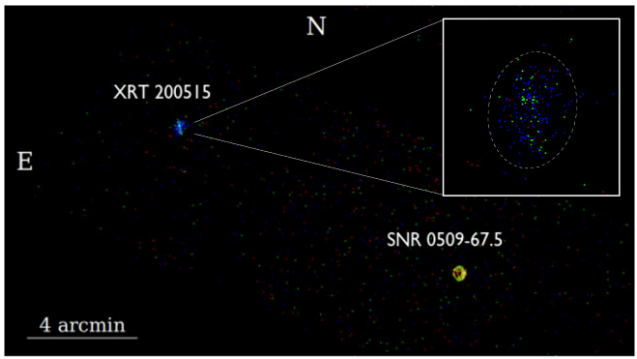

**Figure 13.** ACIS-S image for *Chandra* observation ObsID 23022 showing the target, SNR 0509−67.5, on the bottom right and the transient event, XRT 200515, on the top left. For this image, the event file was filtered to include only the time interval $t_0 + 920$ s around the flare. Red counts correspond to photons in the 0.3–1.2 keV band, green counts correspond to photons in the 1.2–2.4 keV band, and blue photons correspond to photons in the 2.4–7 keV band. The inset image is a 1.0 arcmin × 1.0 arcmin zoomed-in view. The dashed ellipse has semiminor and semimajor axes of 15 arcsec × 20 arcsec, and is the source region used for spectral extraction.

**Table 8.** Properties of the *Chandra* observation ObsID 23022 and source 2CXO J051117.2−672556 associated with XRT 200515.

| ObsID 23022 | |
| --- | --- |
| Observation start time (UTC) | 2020-05-15 11:45:37 |
| Exposure (ks) | 25.06 |
| **XRT 200515 (2CXO J051117.2−672556)** | |
| Flare start time $t_0$ (UTC) | 2020-05-15 18:36:46 |
| RA, Dec. (J2000) | 5:11:17.17−67:25:55.9 |
| 90 per cent position error radius (arcsec) | 2.0 |
| Off-axis angle (arcmin) | 11.75 |
| S/N | 11.64 |

The field was previously covered by four other *Chandra* observations (ObsIDs 776, 7635, 8554, and 23023) with no source detections at the location of 2CXO J051117.2−672556. We estimated model-independent upper limits to the source flux and luminosity with CIAO tool srcflux. In the pre-flare part of ObsID 23022, we obtained a 90 per cent confidence limit of $L_X < 1.0 \times 10^{34}$ erg s$^{-1}$ in the 0.3–7 keV band at the LMC distance of 50 kpc. Stacking the data from all the ObsIDs with non-detections, including the pre-flare part of ObsID 23022, results in a total observed exposure of approximately ~150 ks, and yields a 90 per cent confidence upper limit on the X-ray luminosity is $L_X < 3 \times 10^{33}$ erg s$^{-1}$.

#### 4.3.2 X-ray temporal analysis

We used the CIAO tool dmextract to extract background-subtracted light curves in several energy bands, from the reprocessed event file of *Chandra* ObsID 23022. We defined an elliptical source extraction region, with semiminor and semimajor axes of 15 and 20 arcsec (matching the point-source PSF at the source location); the local background region was chosen in the same ACIS chip, with an area approximately eight times larger.

Fig. 14 shows the 0.3–7 keV background-subtracted light curve of XRT 200515 with a time resolution of 20 s. The light curve is consistent with no source detection at the location of the transient, before the start of the flare at around 23.5 ks into the observation. The few pre-flare counts are consistent with background noise. The light curve exhibits a strong initial spike with a sharp rise of at least four orders of magnitude in <10 s, containing 44 out of all ~180 flare counts. This initial burst is followed by a sudden drop to a ~800 s long pulsating and decaying tail. We estimate a $T_{90} \sim 580$–740 s for the photons observed in the 0.3–7 keV band,[4] depending on the definition of total flare counts.

Fig. 15 shows the light curve of XRT 200515 at a resolution matching the ACIS frame time of 3.2 s, the hardness ratio, and the energy evolution for the time interval $t_0 + 920$ s. The light curve exhibits a spike in the count rate across only 3 bins (with a total of 4, 31, and 9 counts, respectively), hence the burst duration of <10 s. The rise and fall times of the burst are both between 3.2 and 6.4 s. The maximum count rate at the *Chandra* frame time resolution is ∼9.7 counts s⁻¹, acting as the lower bound for the peak count rate of the burst. Those counts are spatially spread over a PSF area of ∼3000 pixels; therefore, pile-up is not an issue. We evaluated the hardness ratio evolution during the flare with the Bayesian estimation method BEHR (Park et al. 2006). Here, the hardness ratio is defined as

$$HR = \frac{h - m - s}{h + m + s}, \qquad (13)$$

where $s$ is the number of soft photons (0.3–1.2 keV), $m$ is the number of medium photons (1.2–2 keV), and $h$ is the number of hard photons (2–7 keV) in each bin. We also track the running average of the photon energies during the flare with a moving window of ±10 counts. The hardness ratio and energy evolution indicate spectral softening during the flare, with the hardness ratio starting at 1.0 during the hard burst peak and decreasing to a range of 0.4–0.6 in the tail, highlighting the notable spectral variability of XRT 200515.

---

[3] https://cxc.harvard.edu/cal/ASPECT/celmon
[4] $T_{90}$ is the time interval during which the cumulative number of counts increases from 5 to 95 per cent of the total flare counts (Kouveliotou et al. 1993).

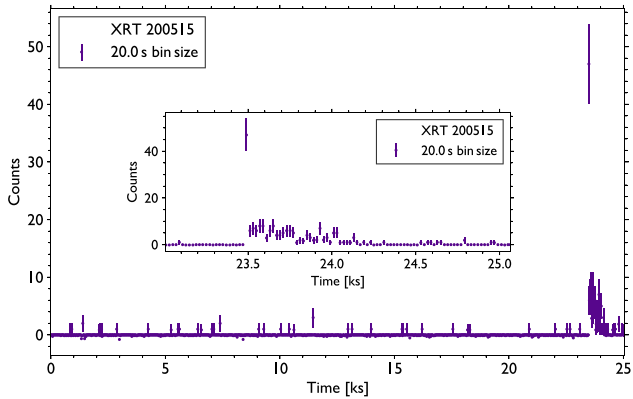

**Figure 14.** Background-subtracted light curve of XRT 200515 in the 0.3–7 keV energy range with a bin size of 20 s. The zero start time is taken as the start of the *Chandra* observation ObsID 23022, and the time interval shown is the full exposure time of 25.06 ks. The inset shows the last ∼2 ks of the observations, and captures the initial burst and tail of the transient event. The bin size is chosen to better visualize the oscillatory decay of the tail. The actual duration of the initial burst peak is <10 s. The presence of a few negative counts arises from the process of background subtraction.

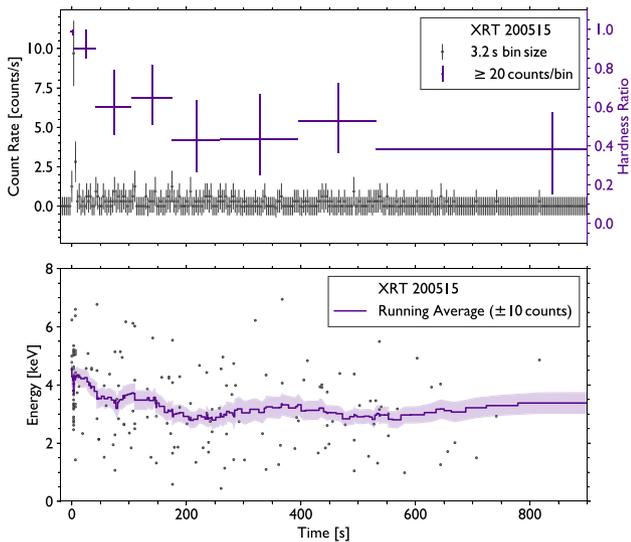

**Figure 15.** *Upper panel:* Background-subtracted count rate light curve of XRT 200515 in the 0.3–7 keV energy range at the full *Chandra* detector resolution of 3.2 s and hardness ratio evolution obtained for a minimum of 20 counts per bin. The zero start time is taken as the flare start time $t_0$ of XRT 200515, and the time interval shown is $t_0 + 920$ s. The initial peak is <10 s long and is very hard, while the ∼800 s long oscillatory tail is significantly softer. *Bottom panel:* The energy evolution during the flare obtained from the running average energy with a moving window of ±10 counts, showing significant spectral variability and softening during the flare. The scatter points represent the time and energy of individual photons from XRT 200515 in the event file associated with *Chandra* observation ObsID 23022.

**Table 9.** Best-fitting parameters of the *Chandra* spectrum of XRT 200515, fitted with Cash statistics, for an absorbed power law and an absorbed blackbody model. Given the relatively low number of counts, parameter uncertainties are reported at the confidence interval $\Delta C = \pm 1.0$: this is asymptotically equivalent to the 68 per cent confidence interval ($1\sigma$) in $\chi^2$ statistics.

| Parameter | Value |
|---|---|
| tbabs × power law | |
| $N_H$ ($10^{22}\,1\,\mathrm{cm}^{-2}$) | $0.58^{+0.45}_{-0.38}$ |
| $\Gamma$ | $0.50^{+0.32}_{-0.31}$ |
| $A_{\mathrm{PL}}$ (photons keV⁻¹ s⁻¹ cm⁻² at 1 keV) | $3.5^{+2.2}_{-1.2} \times 10^{-4}$ |
| Cstat | 132.7 (137 dof) |
| $P_{\mathrm{null}}$ | $3.5 \times 10^{-3}$ |
| tbabs × bb | |
| $N_H$ ($10^{22}\,1\,\mathrm{cm}^{-2}$) | $0.05^{+0.43}_{-0.05}$ |
| $kT_{\mathrm{bb}}$ (keV) | $1.81^{+0.29}_{-0.26}$ |
| $A_{\mathrm{bb}}$ | $1.2^{+0.4}_{-0.2} \times 10^{-4}$ |
| Cstat | 129.6 (137 dof) |
| $P_{\mathrm{null}}$ | $1.2 \times 10^{-2}$ |

### 4.3.3 X-ray spectral analysis

We used the CIAO tool `specextract` to extract the spectrum and the associated response and ancillary response files from the reprocessed event file of *Chandra* ObsID 23022. We used the same source and background extraction regions defined for the light curve extraction. To improve the signal-to-noise ratio of the source, we extracted the spectrum only from the time interval $t_0 + 920$ s. We binned the spectrum to a minimum of 1 count per bin with the `grppha` task within the FTOOLS package suite (Blackburn 1995) from NASA's High Energy Astrophysics Science Research Center (HEASARC).[5] For all spectral modelling and flux estimates, we used the XSPEC software version 12.13.0 (Arnaud 1996). With only 179 net counts, we are unable to fit complex spectral models; thus, we limit our analysis to the simplest one-component models representative of opposite scenarios: a power law (`powerlaw`) and a blackbody model (`bbody`), both modified by photo-electric absorption (`tbabs`). In both cases, we adopted the Tuebingen–Boulder absorption model with Wilms abundances (Wilms, Allen & McCray 2000). We minimized the Cash statistic (Cash 1979). As we do not have enough counts for $\chi^2$ fitting.

The best-fitting power-law model (Table 9 and Fig. 16) has a photon index of $\Gamma = 0.5 \pm 0.3$. The fit statistics yield a null hypothesis probability of $3.5 \times 10^{-3}$, with a Cstat value of 132.7 for 137 degrees of freedom. For the blackbody model, the best-fitting temperature is $kT_{bb} = 1.8 \pm 0.3$ keV (Table 9). The fit statistics yield a null hypothesis probability of $1.2 \times 10^{-2}$, with a Cstat value of 129.6 for 137 degrees of freedom. The reason this blackbody spectrum may appear hard in the *Chandra* band, resembling a $\Gamma \sim 0.5$ power law, is that at a temperature of $kT_{bb} \sim 2$ keV, the ACIS detector samples only the peak and the Rayleigh–Jeans (rising) portion of the blackbody emission. We can use either model to determine an average conversion between the count rate and luminosity. This will then enable us to estimate the peak luminosity in the initial spike, for which we have previously estimated a peak count rate of $\gtrsim 10$ counts s$^{-1}$. The best-fitting power law model implies a peak flux of $F_p \gtrsim 5.6 \times 10^{-10}$ erg s$^{-1}$ cm$^{-2}$, a total flare fluence of $E_f \gtrsim 1.1 \times 10^{-8}$ erg cm$^{-2}$, and a peak unabsorbed 0.3–10 keV luminosity[6] of $L_X \gtrsim 1.7 \times 10^{38}$ erg s$^{-1}$ at the LMC distance of 50 kpc. For the best-fitting blackbody model, the peak flux and flare fluence would be $F_p \gtrsim 4.0 \times 10^{-10}$ erg s$^{-1}$ cm$^{-2}$ and $E_f \gtrsim 0.8 \times 10^{-8}$ erg cm$^{-2}$, respectively. The peak unabsorbed 0.3–10 keV luminosity would be $L_X \gtrsim 1.2 \times 10^{38}$ erg s$^{-1}$ and the peak bolometric luminosity would be $L_{bol} \gtrsim 1.5 \times 10^{38}$ erg s$^{-1}$. These values should be considered conservative lower limits for two reasons: (i) the peak count rate provides only a lower bound estimate, as it is constrained by the *Chandra* frame time resolution of the observations, potentially underestimating the true peak count rate; and (ii) the conversion factor applied is derived from the average spectrum over the entire flare, even though the spectrum of the initial spike is significantly harder compared to the tail, as shown in Fig. 15.

[5] http://heasarc.gsfc.nasa.gov/ftools

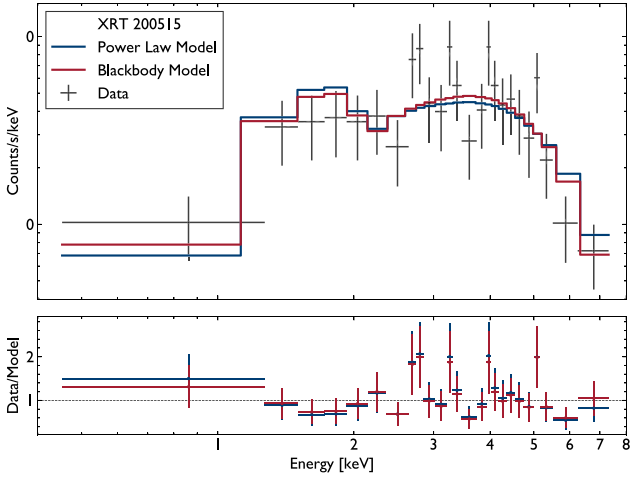

**Figure 16.** *Upper panel:* Observed X-ray spectral energy distribution from counts in the time interval $t_0 + 920$ s around XRT 200515, the best-fitting power-law model (blue) and best-fitting blackbody model (red). The data have been rebinned to a minimum SNR of 2.5 per bin for plotting purposes only; a binning of 1 count per bin was instead used for the fitting (cash statistics). *Lower panel:* Data-to-model ratio between the data and the best-fitting models.

### 4.3.4 High-energy counterpart search

We searched for potential detections of XRT 200515 by other high-energy facilities. However, no significant X-ray or $\gamma$-ray events in the field around the X-ray source coordinates and flare start time $t_0$ reported in Table 8 were detected by the Fermi Gamma-ray Space Telescope (*Fermi*), the Burst Alert Telescope (*BAT*) on the Neil Gehrels Swift Observatory (*Swift*), the International Gamma-Ray Astrophysics Laboratory (*INTEGRAL*), or the Monitor of All-sky X-ray Image (*MAXI*). *LIGO* was not operational during the time of the FXT, hence no gravitational wave signal could have been detected if the origin of XRT 200515 was a compact object merger.

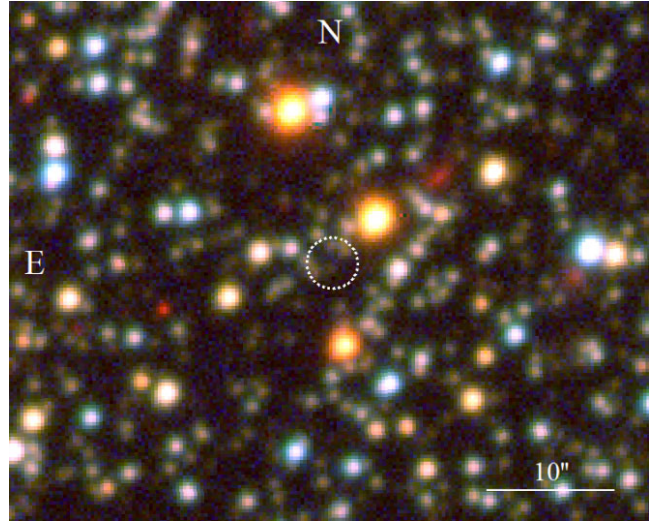

**Figure 17.** *SMASH* survey colour image of the field of XRT 200515 created from the deepest available stacked images in the *u*, *g*, and *i* bands. Red corresponds to the *i* band, green to the *g* band, and blue to the *u* band. The dashed circle has a radius of 2 arcsec, and is the 90 per cent position uncertainty of the *Chandra* source.

### 4.3.5 Optical counterpart search

We used the X-ray source coordinates reported in Table 8 to search for optical counterparts to XRT 200515. The field of XRT 200515 was covered by the Survey of Magellanic Stellar History (*SMASH*; Nidever et al. 2017), a deep optical survey in the *ugriz* bands with the Dark Energy Camera (DECam) mounted on the Víctor M. Blanco Telescope at the Cerro Tololo Inter-American Observatory (CTIO) in Chile. We used the Astro Data Lab Jupyter Notebook server (Nikutta et al. 2020; Juneau et al. 2021) to access and visualize the *SMASH* catalogue.[7] Fig. 17 shows a colour image of the field created from the deepest available stacked images in the *u*, *g*, and *i* bands; the $5\sigma$ detection limits in these bands are 23.9, 24.8, and 24.2 mag, respectively. The images were taken on 2015 December 7 with exposure times of 1179, 981, and 1179 s, respectively. The astrometry of the *SMASH* images is calibrated on the *Gaia* DR3 reference frame, thus their positional uncertainty is negligible compared to the X-ray source position uncertainty. Within the *Chandra* position error circle in Fig. 17, there is no obvious optical counterpart that stands out in brightness or colour from the surrounding stellar population. We performed relative photometry on the sources inside the error circle, comparing them to several nearby sources with known positions and brightnesses listed in the *Gaia* DR3 catalogue. We used the *SMASH g* band as the closest approximation to *Gaia*'s *G* band. We estimate the brightest optical source within the error circle to have a Vega magnitude of $g = 22.7 \pm 0.1$ mag, corresponding to an absolute magnitude of $M_g \approx 4.2$, assuming it is in the LMC. Additionally, three other point-like sources are detected with $g$ band magnitudes in the range of 23–24 mag. All four sources appear point-like, consistent with the seeing conditions of the *SMASH* survey, with no evidence of any spatially extended background galaxies. The three brightest stars visible in Fig. 17 within ∼12 arcsec of the *Chandra* source are solar-mass stars on the red giant branch, indicative of an old stellar population. The lack of bright optical counterparts and the short burst duration of <10 s rules out a stellar flare from a foreground Galactic low-mass star (Güdel 2004; Reale 2007; Reale & Landi 2012; Pye et al. 2015; Kuznetsov & Kolotkov 2021). A flare from a Be/X-ray binary or any other HMXB in the LMC is also excluded by the lack of a bright optical counterpart (Ducci, Mereghetti & Santangelo 2019; Ducci et al. 2022). The temporal and spectral properties of XRT 200515, combined with the absence of an optical counterpart, suggest three possibilities: (i) a relativistic jet phenomenon, such as a $\gamma$-ray burst (GRB); (ii) a rapid, high-energy process linked to extreme magnetic fields, such as a GMF; or (iii) a thermonuclear Type I X-ray burst caused by surface nuclear burning on a neutron star.

---

[6]X-ray luminosities are typically quoted in the standardized energy band 0.3–10 keV to enable comparison of measurements across different observatories.

[7]https://datalab.noirlab.edu/smash/smash.php

### 4.3.6 Gamma-ray burst from a compact object merger?

Evidence in favour or against the association of at least some *Chandra* FXTs with low-luminosity long-GRBs or off-axis short-GRBs (see Berger 2014 for a review), at moderate or high redshifts, is extensively discussed in Quirola-Vásquez et al. (2022), Quirola-Vásquez et al. (2023), and Wichern et al. (2024). A detailed re-investigation of this issue is beyond the scope of this work. Here, we simply point out that XRT 200515, like the other *Chandra* FXTs in the literature, does not have any $\gamma$-ray detection. On the other hand, XRT 200515 has a significantly harder spectrum ($\Gamma = 0.5 \pm 0.3$) in the *Chandra* band than the rest of the FXT sample, all of which have photon indices of $\Gamma > 1$ (Jonker et al. 2013; Glennie et al. 2015; Bauer et al. 2017; Xue et al. 2019; Lin et al. 2022; Quirola-Vásquez et al. 2022; Eappachen et al. 2023; Quirola-Vásquez et al. 2023). A photon index of $\Gamma \sim 0.5$ below 10 keV is indeed expected and observed from both core-collapse GRBs and compact-merger GRBs (Ghirlanda et al. 2009; Bromberg et al. 2013; Oganesyan et al. 2018; Ravasio et al. 2019; Toffano et al. 2021). This might support the association of the initial spike of XRT 200515 with a GRB. However, the presence and properties of the ∼800 s tail (candidate GRB afterglow) is puzzling. The $T_{90} \sim 580$–740 s value for XRT 200515 is significantly shorter than in most other *Chandra* FXTs (Lin et al. 2022; Quirola-Vásquez et al. 2022; Quirola-Vásquez et al. 2023), which have $T_{90}$ values of the order of several ks and are already pushing the limit for a GRB afterglow detection (Wichern et al. 2024). Moreover, XRT 200515's initial burst duration (<10 s), its short rise and fall times (3.2–6.4 s), and the lack of a peak plateau are inconsistent with the light curves of *Chandra* FXTs interpreted as magnetar-powered GRBs as the aftermath of a binary neutron star merger, such as CDF-S XT1 (Bauer et al. 2017), CDF-S XT2 (Xue et al. 2019), and the sample in Lin et al. (2022). Finally, the lack of any optical evidence for a host galaxy is another element disfavouring the high-redshift GRB interpretation.

### 4.3.7 Giant magnetar flare from a soft gamma repeater?

Based on its temporal and spectral variability, it is tempting to interpret XRT 200515 as a rare GMF from an SGR (Mereghetti 2008; Turolla, Zane & Watts 2015) in the LMC or behind it, which can easily explain the burst's strong increase of at least four orders of magnitude in <10 s (Coti Zelati et al. 2018). Similar to XRT 200515,

GMFs are characterized by a short and hard initial spike and a longer and softer, pulsating tail. GMFs are extremely rare, with only a select few ever observed. Well-studied examples are SGR 0526−66 in the LMC (Mazets et al. 1979), and the Galactic sources SGR 1900+14 (Hurley et al. 1999) and SGR 1806−20 (Hurley et al. 2005; Israel et al. 2005; Palmer et al. 2005). More recently, GMFs have been identified in M 31 (Mazets et al. 2008), NGC 253 (Fermi-LAT Collaboration et al. 2021; Roberts et al. 2021; Svinkin et al. 2021; Trigg et al. 2024), and M 82 (Mereghetti et al. 2024). All of these have been observed by high time resolution instruments in the hard X-rays and soft $\gamma$-rays with luminosities above $10^{46}$ erg s$^{-1}$ for a fraction of a second in the initial spike. The tails of GMFs are often modulated by magnetar spin periods of 2–12 s, leading to quasi-periodic oscillations (QPOs). For XRT 200515, there is no hard X-ray or $\gamma$-ray detection, despite the LMC direction being in good visibility for most of the previously mentioned high-energy facilities. We were unable to identify any significant periodicities in the tail of XRT 200515 through periodogram analysis, which is unsurprising given the low time resolution of *Chandra* observations. No X-ray activity has been observed by *Chandra* or other X-ray telescopes in the years before or after XRT 200515, which may be because SGRs are very faint when they are not bursting. The strongest argument against a magnetar in the LMC as the origin of XRT 200515 is that magnetars are short-lived objects ($\lesssim 10^5$ yr) associated with young stellar populations (Olausen & Kaspi 2014; Nakano et al. 2015; Mondal 2021). Even allowing for the persistence of magnetar-like activity in ordinary radio pulsars as old as ∼ $10^7$ yr (Rea et al. 2010), this scenario is still inconsistent with the old stellar population (several Gyr) in the LMC field shown in Fig. 17. The nearest star-forming regions in the LMC are ∼10 arcmin (∼150 pc) away. If (in a very contrived scenario), we assume that XRT 200515 is powered by a young neutron star ejected from one of those regions, we estimate a characteristic time of 1 Myr to travel that distance at a speed of 150 km s$^{-1}$. Therefore, if XRT 200515 is a GMF, it must be located behind the LMC, in a low-redshift galaxy (Hurley et al. 2005; Tanvir et al. 2005). Since GMFs have been observed only a few times and never at soft X-ray energies, their properties in the soft X-ray band detectable by *Chandra* remain largely unexplored. XRT 200515 could indeed be the first GMF detected at soft X-ray energies. Distinguishing distant short GRBs from GMFs has historically been difficult and there are multiple studies suggesting that a subset of short GRBs are actually extragalactic GMFs (Hurley et al. 2005; Palmer et al. 2005; Tanvir et al. 2005; Ofek et al. 2006; Mazets et al. 2008; Hurley 2011; Yang et al. 2020; Svinkin et al. 2021; Negro & Burns 2023). Just as for the distant GRB interpretation, the non-detection of any optical counterpart remains puzzling for a distant GMF scenario, unless we are dealing with a very distant and exceptionally luminous GMF.

### 4.3.8 Thermonuclear X-ray burst from a quiet LMXB in the LMC?

If XRT 200515 is in the LMC, a peak luminosity near the Eddington luminosity $L_{Edd} \sim 10^{38}$ erg s$^{-1}$ and sharp rise time of the flare suggests a Type I X-ray burst interpretation, which is a thermonuclear explosion on the surface of a weakly magnetized, accreting neutron star (Lewin, van Paradijs & Taam 1993; Strohmayer & Bildsten 2003; Galloway et al. 2008, 2020; Galloway & Keek 2021; Alizai et al. 2023). The old stellar population in the field of XRT 200515 is consistent with the presence of neutron star LMXBs. Following the definition of burst time-scale $\tau = E_f/F_p$ in Galloway et al. (2008), we estimate $\tau \sim 20$ s for XRT 200515, which is consistent with Type I X-ray bursts (Galloway & Keek 2021; Alizai et al. 2023). The fitted temperature $kT_{bb} \sim 2$ keV when the average spectrum is fitted with a simple blackbody, and the softening of the spectrum (temperature decrease) in the tail is also typical of Type I X-ray bursts (Galloway et al. 2008; Güver, Psaltis & Özel 2012; Galloway et al. 2020). On the other hand, several observed properties of XRT 200515 are unusual for Type I X-ray bursts. In particular, most Type I X-ray bursts occur when the persistent luminosity (proportional to the accretion rate) of a LMXB is $L_X > 10^{-4} L_{Edd}$ (and, in most cases, $L_X > 10^{-3} L_{Edd}$) (Galloway et al. 2008). Instead, in the initial part of ObsID 23022, the upper limit on the X-ray luminosity at the position of XRT 200515 is $L_X < 10^{-4} L_{Edd}$, so that the X-ray flux increased by at least four orders of magnitudes. On another note, the sharp decline after the initial burst of XRT 200515 would be unusual for Type I X-ray bursts, which typically exhibit a gradual and exponential decay. However, note that most Type I X-ray bursters were observed by the Rossi X-Ray Timing Explorer (*RXTE*; Jahoda et al. 1996), which has a high time resolution. The low time resolution of *Chandra* may have obscured such a decay for XRT 200515. Moreover, most Type I bursts tend to repeat every few hours (Galloway et al. 2008); instead, XRT 200515 is the only event detected at that location over a total observed time of ~150 ks. No LMXB has ever been noted at that position before or after the event. The time interval between bursts is related to an index $\alpha$ defined as the ratio between the integrated persistent fluence between subsequent bursts and the burst fluence; from a comparison of the energy released by accretion (contributing to the persistent fluence) and by thermonuclear burning (burst fluence), we expect $\alpha \gtrsim 40$, in agreement with the observations of Type I bursts (Galloway et al. 2008). If we apply the same criterion ($\alpha \gtrsim 40$) to the persistent and flare fluences of XRT 200515, we would have to wait $> 10^7$ s (4 months) to observe another similar event, assuming the persistent flux level upper limit in ObsID 23022 before the transient event. This waiting time extends to at least one year if we assume the persistent flux upper limit derived from the stacked ~150 ks *Chandra* observations. Only a few one-off bursts from Galactic neutron stars at a very low persistent luminosity ($L_X \sim 10^{32}$–$10^{33}$ erg s$^{-1}$) were found by Cornelisse et al. (2002a, b) and the Magellanic Bridge (Haberl et al. 2023). If XRT 200515 is a Type I X-ray burst, it is the first extragalactic Type I X-ray burster in the LMC and represents the tip of the iceberg for a vast population of faint LMXBs in nearby galaxies, too dim to be detected by *Chandra* or *XMM–Newton*, but which may occasionally reveal themselves via thermonuclear bursts with a long duty cycle.

### 4.3.9 Concluding remarks and outlook for XRT 200515

XRT 200515 is a unique and intriguing extragalactic *Chandra* FXT. The combination of its temporal and spectral properties is unlike any of the other *Chandra* FXT samples. Based on our analysis, the two most likely scenarios for XRT 200515 are (i) a distant GMF from an SGR behind the LMC; the first observed in the low X-ray energy band, missed by any other high-energy facilities, or (ii) an unusual Type I X-ray burst from a previously unknown faint LMXB, the first extragalactic X-ray burster in the LMC. Nevertheless, both of these interpretations come with their own unique challenges. XRT 200515 could, in fact, represent an entirely new type of astronomical phenomenon. After all, the primary objective of our work was to use machine learning to find rare, needle-in-the-haystack anomalies hidden within vast astronomical data sets. We invite further detailed studies of XRT 200515 to evaluate our interpretations and explore alternative scenarios, such as potential associations with a fast radio burst or a supernova shock breakout. We highly recommend follow-up multiband observations at the source coordinates of XRT 200515 to better constrain its nature. Lastly, we note that XRT 200515 and the second transient discovered by Glennie et al. (2015), XRT 120830, have remarkably similar temporal evolutions in their light curves

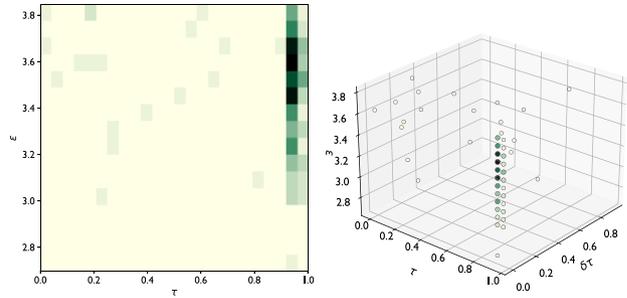

**Figure 18.** $E - t$ maps event file representation (left) and $E - t - dt$ cubes event file representation (right) of XRT 200515 in *Chandra* observation ObsID 23022. The catalogue identifier for XRT 200515 is `CATALOG_ID: 23022_122`.

(J. Irwin, personal communication, November 2024); however, with very different spectral properties ($\Gamma \sim 2.5$ for XRT 120830 versus $\Gamma \sim 0.5$ for XRT 200515). We leave a detailed comparative analysis of these transients for future work.

Fig. 18 shows the $E - t$ maps and $E - t - dt$ cubes event file representations for XRT 200515. These exhibit high counts at high energies in a narrow time window, which is in line with the hard spectrum and transient nature of XRT 200515.

### 4.4 Technical caveats

The main technical caveat of our approach is related to the representation of event files. While our new event file representations enable a simple, yet powerful representation learning approach to find new and rare X-ray transients, any simplification of raw event files, like the fixed number of time bins we use across all event files, is associated with a loss of information. This could lead to us missing a small amount of transients. To minimize this, we have implemented a rigorous approach to justify the resolution of the event file representations in Section 3.1. Moreover, flares, in particular known extragalactic FXTs, cluster notably well in our representation spaces. This is because their distinctive features are less dependent on the temporal binning resolution in the $E - t$ maps and $E - t - dt$ cubes. To improve the effectiveness of dip searches with our proposed method, we suggest using higher resolution event file representations. Nevertheless, our comprehensive transient candidate catalogue includes numerous newly identified transients that were previously overlooked by other X-ray transient searches in the *Chandra* archive. Among these is the remarkable needle-in-the-haystack event XRT 200515 discovered in this work, underscoring the effectiveness of our method. A follow-up representation learning algorithm will learn informative features from raw and unbinned event files while accounting for the Poisson nature of X-ray observations (Song et al., in preparation).

## 5 CONCLUSION

We have introduced a novel representation learning method, the first of its kind applied to X-ray event files, enabling downstream tasks such as anomaly detection, similarity searches and unsupervised classification high-energy astrophysics. We have used the learned representation to investigate time-domain properties of sources in the *Chandra* archive, with a particular emphasis on the discovery of X-ray transients. As a result, we have compiled the identified X-ray flares and dips in a comprehensive catalogue of transient candidates. Notably, our method led to the discovery of XRT 200515; a previously unidentified extragalactic FXT with unique temporal and spectral properties, representing a genuine needle-in-the-haystack discovery. Our key results are as follows:

(i) We introduce novel event file representations, the $E - t$ maps and $E - t - $ d$t$ cubes, which capture both temporal and spectral information.

(ii) We apply two feature extraction methods to the event file representations, PCA and sparse autoencoder neural networks, to extract or learn informative features that can be utilized for downstream tasks, such as anomaly detection, similarity searches and unsupervised classification.

(iii) We project the learned features to two-dimensional embedding spaces, enabling interpretable queries of analogues to objects of interest based on their temporal and spectral properties.

(iv) We cluster the embedding spaces with DBSCAN, successfully isolating previously identified X-ray transients. We identify new transient candidates within specific transient-dominant clusters or through nearest-neighbour searches using kNN.

(v) We compile a catalogue of the X-ray transient candidates, including 3447 flares and 112 dips, and make it openly accessible to the community and the broader scientific audience.

(vi) We report the discovery of XRT 200515, a rare extragalactic FXT characterized by unique temporal and spectral features. We explore its potential origins and suggest that it may be associated with one of the following scenarios, presented in no particular order:

    (a) A rare GMF from an SGR behind the LMC, marking the first GMF detected in the low X-ray energy range covered by telescopes like *Chandra*, *XMM–Newton*, *Swift-XRT*, *eROSITA*, or *Einstein Probe*.

    (b) A rare extragalactic Type I X-ray burst from a faint LMXB in the LMC, representing the first such detection in the LMC.

    (c) A new type of astronomical phenomenon and a genuine anomaly, previously hidden in the vast *Chandra* archive.

XRT 200515 was only detected by *Chandra*, with no identified optical counterparts. We strongly encourage a multiwavelength search for additional signals from the source associated with XRT 200515 to better understand its origin and nature.

Our work advances time-domain high-energy astrophysics by making the *Chandra* transient candidates catalogue[8] (Dillmann et al. 2025) publicly available and open-sourcing the representation learning-based transient search pipeline.[9] The catalogue enables queries to identify and characterize new *Chandra* transients. Future work involves using the learned representations for different downstream tasks, experimenting with different event file representations, applying the detection pipeline to additional high-energy archives, and adapting it to a variety of other scientific data sets, paving the way for further machine learning driven discoveries of rare transients and other scientific anomalies.

## ACKNOWLEDGEMENTS

This research has made use of data obtained from the CSC, and software provided by the Chandra X-ray Center (CXC) in the CIAO application package.

Steven Dillmann's work was partially funded by the UK government's Turing Scheme and mainly carried out at the Center for

Astrophysics | Harvard & Smithsonian as part of the SAO Predoctoral Program, with the support of AstroAI. SD acknowledges hospitality at the Institute of Astronomy at the University of Cambridge and at the Stanford Center for Decoding the Universe (Stanford Data Science) during the later parts of this project. Roberto Soria's work was performed in part at the Aspen Center for Physics, which is supported by National Science Foundation grant PHY-2210492. RS acknowledges support and hospitality at the National Astronomical Observatories of China (NAOC) in Beijing, during part of this project; he also acknowledges support from the INAF grant number 1.05.23.04.04.

We thank Edo Berger, Massimiliano De Pasquale, Ken Ebisawa, Duncan Galloway, Jimmy Irwin, Peter Jonker, Daniel Kocevski, Amy Lien, Sandro Mereghetti, Daniel Muthukrishna, Nicola Omodei, Jonathan Quirola-Vásquez, Shivam Raval, Ashley Villar, and Silvia Zane for their fruitful discussions.

## DATA AVAILABILITY

The data used in this paper, composed of X-ray event files and source detection regions, was obtained from the publicly available CSC, using their public interfaces (https://cxc.cfa.harvard.edu/csc/). The catalogue of transient candidates and the clustered embedding spaces generated using our unsupervised representation learning method can be accessed in the supplementary material. The catalogue and lightcurve images for each flare and dip candidate are also available at: https://zenodo.org/records/14589318. All intermediate data products, i.e. as $E - t$ maps, $E - t - $ d$t$ cubes, principal components and latent features, feature embeddings and embedding clusters can be produced using the code provided in the GitHub repository https://github.com/StevenDillmann/ml-xraytransients-mnras.

## REFERENCES

Abadi M. et al., 2016, Proc. 12th USENIX Conference on Operating Systems Design and Implementation. USENIX Association, Savannah, GA, p. 265
Alizai K. et al., 2023, MNRAS, 521, 3608
Alp D., Larsson J., 2020, ApJ, 896, 39
Arcodia R. et al., 2021, Nature, 592, 704
Arnaud K., 1996, in Jacoby G. H., Barnes J., eds, ASP Conf. Ser. Vol. 101, Astronomical Data Analysis Software and Systems V. Astron. Soc. Pac., San Francisco, p. 17
Astropy Collaboration, 2013, A&A, 558, A33
Bauer F. E. et al., 2017, MNRAS, 467, 4841
Bellm E. C. et al., 2019, PASP, 131, 018002
Bengio Y., Courville A., Vincent P., 2013, IEEE Trans. Pattern Anal. Mach. Intell., 35, 1798
Berger E., 2014, ARA&A, 52, 43
Blackburn J. K., 1995, in Shaw R. A., Payne H. E., Hayes J. J. E., eds, ASP Conf. Ser. Vol. 77, Astronomical Data Analysis Software and Systems IV. Astron. Soc. Pac., San Francisco, p. 367
Bromberg O., Nakar E., Piran T., Sari R., 2013, ApJ, 764, 179
Burrows D. N. et al., 2005, Space Sci. Rev., 120, 165
Caliński T., Harabasz J., 1974, Commun. Stat.-Theor. Methods, 3, 1
Cash W., 1979, ApJ, 228, 939
Chakraborty J., Kara E., Masterson M., Giustini M., Miniutti G., Saxton R., 2021, ApJ, 921, L40
Chambers K. C. et al., 2016, preprint (arXiv:1612.05560)
Colbert E. J. M., Ptak A. F., 2002, ApJS, 143, 25
Cornelisse R. et al., 2002a, A&A, 392, 885
Cornelisse R., Verbunt F., in't Zand J. J. M., Kuulkers E., Heise J., 2002b, A&A, 392, 931
Coti Zelati F., Rea N., Pons J. A., Campana S., Esposito P., 2018, MNRAS, 474, 961

Cover T., Hart P., 1967, IEEE T. Inf. Theor., 13, 21
D'Aì A., Iaria R., Di Salvo T., Riggio A., Burderi L., Robba N. R., 2014, A&A, 564, A62
D'Orazio D. J., Di Stefano R., 2018, MNRAS, 474, 2975
D'Orazio D. J., Di Stefano R., 2020, MNRAS, 491, 1506
Dai Z. G., Wang X. Y., Wu X. F., Zhang B., 2006, Science, 311, 1127
Davies D. L., Bouldin D. W., 1979, IEEE Trans. Pattern Anal. Mach. Intell., PAMI-1, 224
Davis J. E., 2001, ApJ, 562, 575
Deneva J. S. et al., 2016, ApJ, 823, 105
Dey A. et al., 2019, AJ, 157, 168
Di Stefano R., Kong A. K. H., 2004, ApJ, 609, 710
Di Stefano R., Berndtsson J., Urquhart R., Soria R., Kashyap V. L., Carmichael T. W., Imara N., 2021, Nat. Astron., 5, 1297
Dillmann S., Martínez-Galarza R., 2023, Master's thesis, Imperial College London
Dillmann S., Martínez-Galarza J. R., Soria R., Di Stefano R., Kashyap V., 2025, Representation learning for time-domain high-energy astrophysics: Transient candidates catalog of X-ray flares and dips. Zenodo, available at: https://zenodo.org/records/14589318
Ducci L., Mereghetti S., Santangelo A., 2019, ApJ, 881, L17
Ducci L. et al., 2022, A&A, 661, A22
Duchi J., Hazan E., Singer Y., 2011, J. Mach. Learn. Res., 12, 2121
Eappachen D. et al., 2023, ApJ, 948, 91
Evans I. N. et al., 2024, ApJS, 274, 22
Fermi-LAT Collaboration, 2021, Nat. Astron., 5, 385
Freedman D., Diaconis P., 1981, Probability Theory and Related Fields, 57, 453
Fruscione A. et al., 2006, in Silva D. R., Doxsey R. E., eds, Proc. SPIE Conf. Ser. Vol. 6270, Observatory Operations: Strategies, Processes, and Systems. SPIE, Bellingham, p. 62701V
Gaia Collaboration, 2021, A&A, 649, A1
Galloway D. K., Keek L., 2021, in Belloni T. M., Méndez M., Zhang C., eds, Astrophysics and Space Science Library Vol. 461, Timing Neutron Stars: Pulsations, Oscillations and Explosions. Springer, Berlin, Heidelberg, p. 209
Galloway D. K., Muno M. P., Hartman J. M., Psaltis D., Chakrabarty D., 2008, ApJS, 179, 360
Galloway D. K. et al., 2020, ApJS, 249, 32
Ghirlanda G., Nava L., Ghisellini G., Celotti A., Firmani C., 2009, A&A, 496, 585
Glennie A., Jonker P. G., Fender R. P., Nagayama T., Pretorius M. L., 2015, MNRAS, 450, 3765
Green P. J., Kim D. W., Martínez-Galarza R., Yang Q., D'Abrusco R., Rots A., Evans I., the CXC, 2023, available at: https://cxc.cfa.harvard.edu/csc/files/README_CSC2.1p_OIR_SDSSspecmatch.pdf (Accessed October 2023)
Gregory P. C., Loredo T. J., 1992, ApJ, 398, 146
Güdel M., 2004, A&AR, 12, 71
Guest B. T., Borkowski K. J., Ghavamian P., Petre R., Reynolds S. P., Seitenzahl I. R., Williams B. J., 2022, AJ, 164, 231
Güver T., Psaltis D., Özel F., 2012, ApJ, 747, 76
Haberl F. et al., 2023, A&A, 669, A66
Hartigan J. A., Wong M. A., 1979, J. R. Stat. Soc., 28, 100
Hayat M. A., Stein G., Harrington P., Lukić Z., Mustafa M., 2021, ApJ, 911, L33
Hinton G. E., Salakhutdinov R. R., 2006, Science, 313, 504
Hu B. X., D'Orazio D. J., Haiman Z., Smith K. L., Snios B., Charisi M., Di Stefano R., 2020, MNRAS, 495, 4061
Hurley K., 2011, Adv. Space Res., 47, 1337
Hurley K. et al., 1999, Nature, 397, 41
Hurley K. et al., 2005, Nature, 434, 1098
Israel G. L. et al., 2005, ApJ, 628, L53
Ivezić Ž. et al., 2019, ApJ, 873, 111
Jahoda K., Swank J. H., Giles A. B., Stark M. J., Strohmayer T., Zhang W., Morgan E. H., 1996, in Siegmund O. H., Gummin M. A., eds, Proc. SPIE Conf. Ser. Vol. 2808, EUV, X-Ray, and Gamma-Ray Instrumentation for Astronomy VII. SPIE, Bellingham, p. 59

Jansen F. et al., 2001, A&A, 365, L1
Jolliffe I. T., 2002, Principal Component Analysis for Special Types of Data. Springer, New York, p. 338
Jonker P. G. et al., 2013, ApJ, 779, 14
Juneau S., Olsen K., Nikutta R., Jacques A., Bailey S., 2021, Comput. Sci. Eng., 23, 15
Kingma D. P., Ba J., 2014, CoRR, abs/1412.6980, available at: https://api.semanticscholar.org/CorpusID:6628106
Kouveliotou C., Meegan C. A., Fishman G. J., Bhat N. P., Briggs M. S., Koshut T. M., Paciesas W. S., Pendleton G. N., 1993, ApJ, 413, L101
Kouveliotou C., Patel S., Tennant A., Woods P., Finger M., Wachter S., 2003, IAU Circ., 8109, 2
Kovačević M., Pasquato M., Marelli M., De Luca A., Salvaterra R., Belfiore A., 2022, A&A, 659, A66
Kullback S., Leibler R. A., 1951, Ann. Math. Stat., 22, 79
Kuznetsov A. A., Kolotkov D. Y., 2021, ApJ, 912, 81
LeCun Y., Bottou L., Bengio Y., Haffner P., 1998, Proc. IEEE, 86, 2278
Lewin W. H. G., van Paradijs J., Taam R. E., 1993, Space Sci. Rev., 62, 223
Lin D., Webb N. A., Barret D., 2012, ApJ, 756, 27
Lin D., Webb N. A., Barret D., 2014, ApJ, 780, 39
Lin D., Irwin J. A., Berger E., 2021, Astron. Telegram, 14599, 1
Lin D., Irwin J. A., Berger E., Nguyen R., 2022, ApJ, 927, 211
Maas A. L., 2013, in Proc. ICML, p. 3, available at: https://api.semanticscholar.org/CorpusID:16489696
Maaten L. v. d., Hinton G., 2008, J. Mach. Learn. Res., 9, 2579
Mahabal A., Sheth K., Gieseke F., Pai A., Djorgovski S. G., Drake A., Graham M., the CSS/CRTS/PTF Collaboration, 2017, IEEE Symposium Series on Computational Intelligence (SSCI). IEEE, p. 1
Mahalanobis P. C., 1936, in Proc. National Institute of Science of India, On the Generalized Distance in Statistics. India, p. 49
Marshall F. E., Gotthelf E. V., Middleditch J., Wang Q. D., Zhang W., 2004, ApJ, 603, 682
Martínez-Galarza R., Makinen T. L., 2022, in Adler D. S., Seaman R. L., Benn C. R., eds, Proc. SPIE Conf. Ser. Vol. 12186, Observatory Operations: Strategies, Processes, and Systems IX. SPIE, Bellingham, p. 121860J
Martínez-Galarza J. R., Bianco F. B., Crake D., Tirumala K., Mahabal A. A., Graham M., Giles D., 2021, MNRAS, 508, 5734
Masci J., Meier U., Cireşan D., Schmidhuber J., 2011, in Artificial Neural Networks and Machine Learning – ICANN 2011. Springer, Berlin, Heidelberg, p. 52
Mazets E. P., Golentskii S. V., Ilinskii V. N., Aptekar R. L., Guryan I. A., 1979, Nature, 282, 587
Mazets E. P. et al., 2008, ApJ, 680, 545
Mereghetti S., 2008, A&AR, 15, 225
Mereghetti S. et al., 2024, Nature, 629, 58
Metzger B. D., Quataert E., Thompson T. A., 2008, MNRAS, 385, 1455
Mishra-Sharma S., Song Y., Thaler J., 2024, preprint (arXiv:2403.08851)
Modjaz M. et al., 2009, ApJ, 702, 226
Mohale K., Lochner M., 2024, MNRAS, 530, 1274
Mondal T., 2021, ApJ, 913, L12
Muthukrishna D., Mandel K. S., Lochner M., Webb S., Narayan G., 2022, MNRAS, 517, 393
Nair V., Hinton G. E., 2010, in Proc. 27th International Conference on Machine Learning (ICML-10). Omnipress, Haifa, Israel, p. 807
Nakano T., Murakami H., Makishima K., Hiraga J. S., Uchiyama H., Kaneda H., Enoto T., 2015, PASJ, 67, 9
Naul B., Bloom J. S., Pérez F., van der Walt S., 2018, Nat. Astron., 2, 151
Negro M., Burns E., 2023, in Troja E., Baring M. G., eds, IAU Symp. 363, Neutron Star Astrophysics at the Crossroads: Magnetars and the Multimessenger Revolution. p. 284
Ng A. et al., 2011, CS294A Lecture notes, 72, 1
Nidever D. L. et al., 2017, AJ, 154, 199
Nikutta R., Fitzpatrick M., Scott A., Weaver B. A., 2020, Astron. Comput., 33, 100411
Novara G. et al., 2020, ApJ, 898, 37
Ofek E. O. et al., 2006, ApJ, 652, 507
Oganesyan G., Nava L., Ghirlanda G., Celotti A., 2018, A&A, 616, A138
Olausen S. A., Kaspi V. M., 2014, ApJS, 212, 6

Palmer D. M. et al., 2005, Nature, 434, 1107

Park T., Kashyap V. L., Siemiginowska A., van Dyk D. A., Zezas A., Heinke C., Wargelin B. J., 2006, ApJ, 652, 610

Parker L. et al., 2024, MNRAS, 531, 4990

Parmar A. N., White N. E., Giommi P., Gottwald M., 1986, ApJ, 308, 199

Pastor-Marazuela I., Webb N. A., Wojtowicz D. T., van Leeuwen J., 2020, A&A, 640, A124

Patel S. K. et al., 2004, ApJ, 602, L45

Patel S. K. et al., 2007, ApJ, 657, 994

Pearson K., 1901, London, Edinburgh, Dublin Phil. J. Sci., 2, 559

Pedregosa F., 2011, J. Mach. Learn. Res., 12, 2825

Pérez-Díaz V. S., Martínez-Galarza J. R., Caicedo A., D'Abrusco R., 2024, MNRAS, 528, 4852

Predehl P. et al., 2021, A&A, 647, A1

Pye J. P., Rosen S., Fyfe D., Schröder A. C., 2015, A&A, 581, A28

Quirola-Vásquez J. et al., 2022, A&A, 663, A168

Quirola-Vásquez J. et al., 2023, A&A, 675, A44

Radford A. et al., 2021, in Meila M., Zhang T., eds, Proc. 38th International Conference on Machine Learning. PMLR, p. 8748, available at: https://proceedings.mlr.press/v139/radford21a.html( Accessed April 2023)

Ravasio M. E., Ghirlanda G., Nava L., Ghisellini G., 2019, A&A, 625, A60

Rea N. et al., 2010, Science, 330, 944

Reale F., 2007, A&A, 471, 271

Reale F., Landi E., 2012, A&A, 543, A90

Reynolds C. S. et al., 2024, in den Herder J.-W. A., Nikzad S., Nakazawa K., eds, Proc. SPIE Conf. Ser. Vol. 13093, Space Telescopes and Instrumentation 2024: Ultraviolet to Gamma Ray. SPIE, Bellingham, p. 1309328

Rizhko M., Bloom J. S., 2024, preprint (arXiv:2411.08842)

Roberts O. J. et al., 2021, Nature, 589, 207

Scargle J. D., Norris J. P., Jackson B., Chiang J., 2013, ApJ, 764, 167

Simonyan K., Zisserman A., 2014, CoRR, abs/1409.1556, available at: https://api.semanticscholar.org/CorpusID:14124313

Skrutskie M. F. et al., 2006, AJ, 131, 1163

Slijepcevic I. V., Scaife A. M. M., Walmsley M., Bowles M., Wong O. I., Shabala S. S., White S. V., 2024, RAS Techniques and Instruments, 3, 19

Slipski M., Kleinböhl A, Dillmann R., Kass D. M., Reimuller J., Wronkiewicz M., Doran G., 2024, Icarus, 419, 115777

Soderberg A. M. et al., 2008, Nature, 453, 469

Spearman C., 1904, Am. J. Psychol., 15, 72

Strohmayer T., Bildsten L., 2003, preprint (arXiv:astro-ph/0301544)

Sun H., Zhang B., Gao H., 2017, ApJ, 835, 7

Sutskever I., Martens J., Dahl G., Hinton G., 2013, in Proc. 30th International Conference on International Conference on Machine Learning, Vol. 28. p. III–1139

Svinkin D. et al., 2021, Nature, 589, 211

Swartz D. A., Ghosh K. K., Tennant A. F., Wu K., 2004, ApJS, 154, 519

Tanvir N. R., Chapman R., Levan A. J., Priddey R. S., 2005, Nature, 438, 991

The Multimodal Universe Collaboration, 2024, preprint (arXiv:2412.02527)

Toffano M., Ghirlanda G., Nava L., Ghisellini G., Ravasio M. E., Oganesyan G., 2021, A&A, 652, A123

Triga A. C. et al., 2024, A&A, 687, A173

Turolla R., Zane S., Watts A. L., 2015, Rep. Prog. Phys., 78, 116901

Villar V. A., Cranmer M., Berger E., Contardo G., Ho S., Hosseinzadeh G., Lin J. Y.-Y., 2021, ApJS, 255, 24

Walmsley M. et al., 2022, MNRAS, 513, 1581

Weisskopf M. C., Tananbaum H. D., Van Speybroeck L. P., O'Dell S. L., 2000, in Truemper J. E., Aschenbach B., eds, Proc. SPIE Conf. Ser. Vol. 4012, X-Ray Optics, Instruments, and Missions III. SPIE, Bellingham, p. 2

Wenger M. et al., 2000, A&AS, 143, 9

Wichern H. C. I., Ravasio M. E., Jonker P. G., Quirola-Vásquez J. A., Levan A. J., Bauer F. E., Kann D. A., 2024, A&A, 690, A101

Wilms J., Allen A., McCray R., 2000, ApJ, 542, 914

Wisotzki L., Christlieb N., Liu M. C., Maza J., Morgan N. D., Schechter P. L., 1999, A&A, 348, L41

Xue Y. Q. et al., 2019, Nature, 568, 198

Yang G., Brandt W. N., Zhu S. F., Bauer F. E., Luo B., Xue Y. Q., Zheng X. C., 2019, MNRAS, 487, 4721

Yang J. et al., 2020, ApJ, 899, 106

Yang H., Hare J., Kargaltsev O., Volkov I., Chen S., Rangelov B., 2022, ApJ, 941, 104

Yuan W., Zhang C., Chen Y., Ling Z., 2022, in Bambi C., Sangangelo A., eds, Handbook of X-ray and Gamma-ray Astrophysics. Springer Nature, Singapore, p. 86

Zhang B., 2013, ApJ, 763, L22

Zhang G., Helfer T., Gagliano A. T., Mishra-Sharma S., Villar V. A., 2024, preprint (arXiv:2408.16829)

## SUPPORTING INFORMATION

Supplementary data are available at *MNRAS* online.

**suppl_data**

Please note: Oxford University Press is not responsible for the content or functionality of any supporting materials supplied by the authors. Any queries (other than missing material) should be directed to the corresponding author for the article.

## APPENDIX A: EVENT FILE LENGTHS AND DURATIONS

Fig. A1 shows the distribution of the length $N$ and duration $T$ of event files in the data set used in this work.

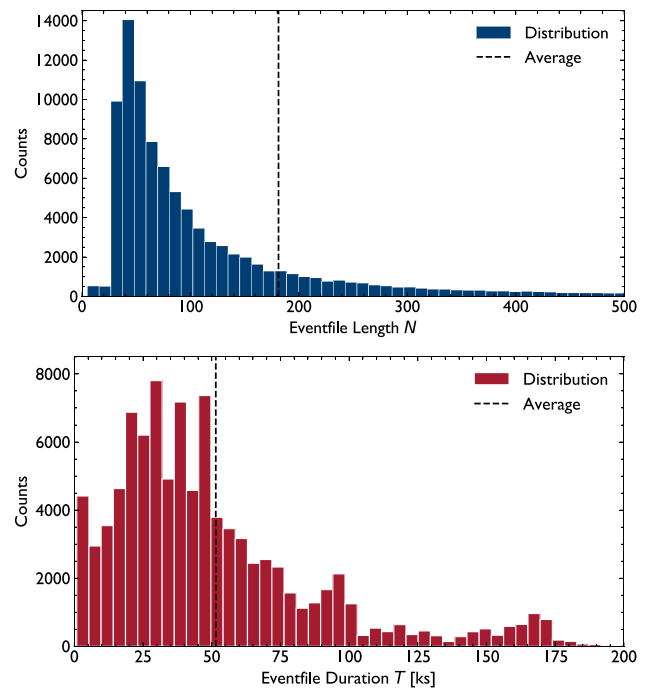

**Figure A1.** Distribution of *Chandra* event file lengths $N$ (top) and durations $T$ (bottom) in the data set used in this work.

## APPENDIX B: AUTOENCODER TRAINING PROCESS

Fig. B1 shows the training process of the autoencoders used in this work.

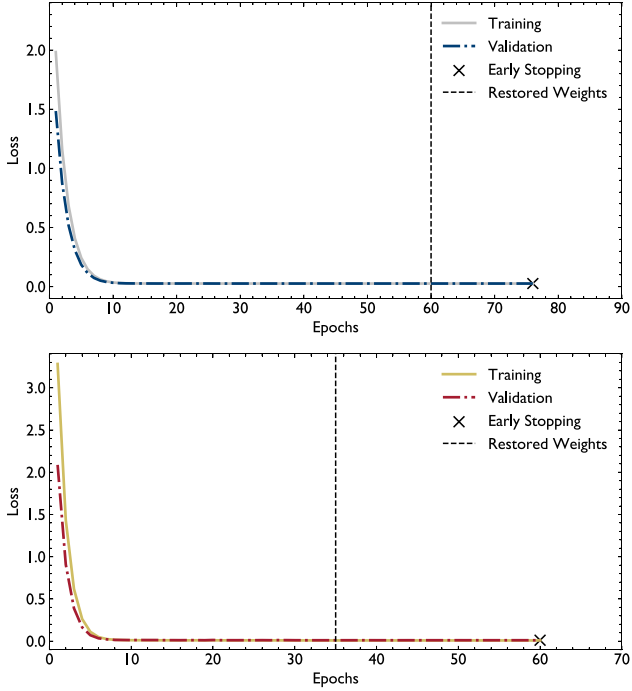

**Figure B1.** Training process for the convolutional autoencoder applied to the $E - t$ maps (top) and fully connected autoencoder applied to the $E - t - \mathrm{d}t$ cubes (bottom). The plots show the evolution of the training and validation loss with the number of epochs.

# APPENDIX C: HYPERPARAMETER OPTIMIZATION

Below, we summarize the optimization strategy for the t-SNE and DBSCAN hyperparameters. For even more details on this approach, please refer to Dillmann & Martínez-Galarza (2023).

## C1 t-SNE hyperparameters

The choice of the `perplexity` and `learning_rate` can have a large impact on the resulting t-SNE embedding space. Ideally, we want the two-dimensional embedding space to effectively capture both energy information (hardness ratio HR) and variability information (variability probability $p_{var}$). Therefore, event files with similar values for HR and $p_{var}$ should live close to each other in the embedding space. We can use this information to define a performance metric for different t-SNE hyperparameter inputs. First, we compute the pairwise distance matrix $\mathbf{D_Z}$ of size $(m, m)$, where the distance $D_{Z_{ij}}$ between points $i$ and $j$ is computed using a Euclidean distance metric. Next, we define the property vector $\mathbf{Y}$, which includes seven CSC properties (hardness ratios $HR_{hm}$, $HR_{hs}$, $HR_{ms}$ and variability probabilities $p_{var}^b$, $p_{var}^h$, $p_{var}^m$, $p_{var}^s$) for each event file. As a measure of similarity between the properties of different points, we can again compute a pairwise similarity matrix $\mathbf{D_Y}$ of size $(m, m)$. To compute the similarity distance $D_{Y_{ij}}$ between sample $i$ and $j$, we use the Mahalanobis distance metric (Mahalanobis 1936). Unlike the Euclidean distance metric, the Mahalanobis distance metric accounts for the correlation between different labels by taking into account the covariance structure of the data. Note that our hardness ratios are correlated with each other, and that the same holds for the variability probabilities. Accounting for these correlations provides a more accurate measure of the similarity distance between different samples. Having computed $\mathbf{D_Z}$ and $\mathbf{D_Y}$, we can define a performance metric that allows us to compare the performance of different t-SNE hyperparameters. The smaller the distance $D_{Z_{ij}}$ between two points $i$ and $j$ in the t-SNE embedding, the smaller should be difference in their associated labels as measured by the distance $D_{Y_{ij}}$. We can thus define a performance metric based on the statistical correlation of $\mathbf{D_Z}$ and $\mathbf{D_Y}$ using the Spearman's rank correlation coefficient $\rho_{ZY}$ (Spearman 1904). The higher $\rho_{ZY}$, the higher is the positive correlation between $\mathbf{D_Z}$ and $\mathbf{D_Y}$ and the better the performance of the t-SNE embedding. The hyperparameter space is given by the ranges `learning_rate` $\in (20, 200)$ with a step size of 20 and `perplexity` $\in (10, 100)$ with a step size of 10. This optimization process is performed using a reduced data set of 15 353 samples for 2000 iterations per hyperparameter combination due to computational constraints. While subsampling, the overall structure of the data was preserved by selecting the same distributions between any combinations of hard, medium, soft, variable and non-variable samples. This ensures that the sample set is representative of the original data. We choose the hyperparameter combination that produces the highest value of $\rho_{ZY}$.

## C2 DBSCAN hyperparameters

Different hyperparameter combinations of `eps` and `minPts` can have a large impact on the resulting DBSCAN clusters. We use a combination of the Davies–Bouldin index DB (Davies & Bouldin 1979) and Calinski–Harabasz index CH (Caliński & Harabasz 1974) as a performance metric to find the optimal DBSCAN hyperparameter inputs. The DB index is a measure of the average similarity between each cluster and its most similar cluster, relative to the average distance between points within each cluster. The DB index is given by the following formula:

$$\mathrm{DB} = \frac{1}{n_c} \sum_{i=1}^{n_c} \max_{j \neq i} \left( \frac{W_i + W_j}{d(c_i, c_j)} \right), \tag{C1}$$

where $n_c$ is the number of clusters, $W_i$ and $W_j$ are the within-cluster sum of squares for cluster $i$ and $j$, and $d(c_i, c_j)$ is the distance between the centroids of clusters $i$ and $j$. On the other hand, the CH index is based on the concept that good clusters should have high intra-cluster similarity (cohesion) measured by the between-cluster dispersion $B$ and low inter-cluster similarity (separation) measured by the within-cluster dispersion $W$. $B$ is the sum of the pairwise distances between cluster centroids, and $W$ is the sum of the pairwise distances between points within each cluster. The CH index is given by the following formula:

$$\mathrm{CH} = \frac{B}{W} \times \frac{m - n_c}{n_c - 1}, \tag{C2}$$

where the scaling factor $\frac{m - n_c}{n_c - 1}$ accounts for the total number of data points $m$ and the number of clusters $n_c$. A lower DB index and higher CH index indicate that the clustering algorithm is more effective in grouping similar data points together and separating different data points into distinct clusters. We thus define the performance metric $\rho_{DC}$ as the ratio of the normalized indices $\mathrm{DB}_n = \frac{\mathrm{DB}}{\max(\mathrm{DB})}$ and $\mathrm{CH}_n = \frac{\mathrm{CH}}{\max(\mathrm{CH})}$ in the hyperparameter space given by `eps` $\in (1.0, 3.0)$ with a step size of 0.1 and `minPts` $\in (10, 30)$ with a step size of 1:

$$\rho_{DBSCAN} = \frac{\mathrm{CH}_n}{\mathrm{DB}_n}. \tag{C3}$$

We choose the hyperparameter combination that produces the highest value of $\rho_{DBSCAN}$.

## APPENDIX D: EMBEDDINGS

Figs D1–D3 show the *2D-PCA* and *2D-AE* embeddings.

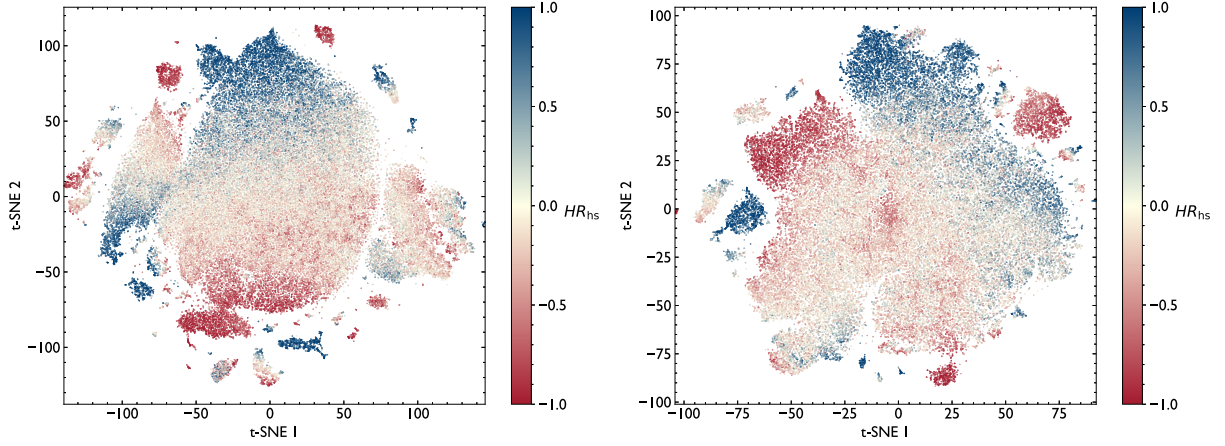

**Figure D1.** Embedding space colour-coded by $HR_{hs}$ for the *2D-PCA* case (left) and *2D-AE* case (right).

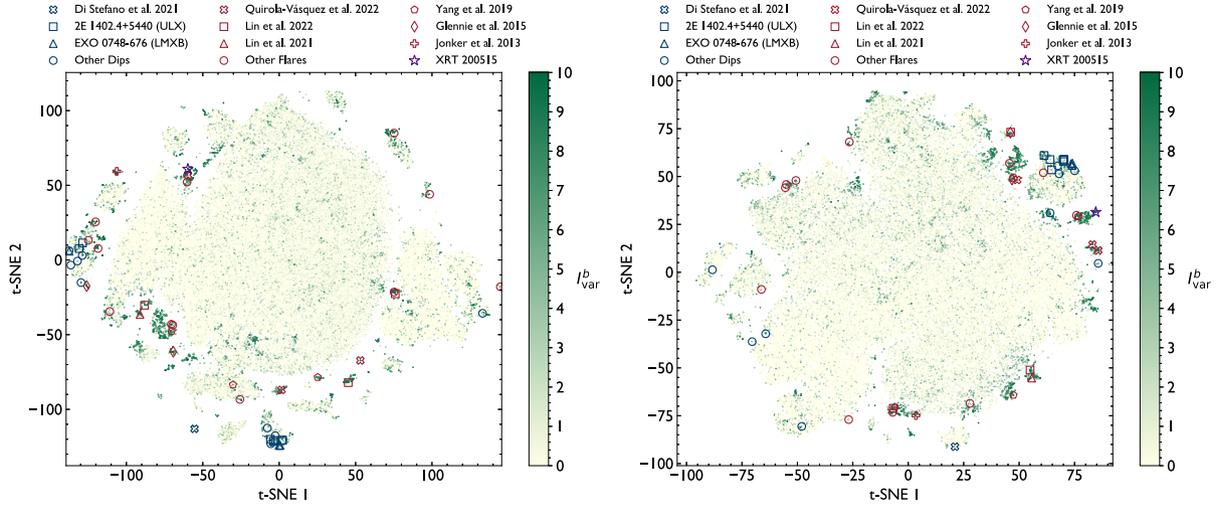

**Figure D2.** Embedding space colour-coded by $I^b_{var}$ for *2D-PCA* (left) and *2D-AE* (right). The bona-fide transients and XRT 200515 are highlighted.

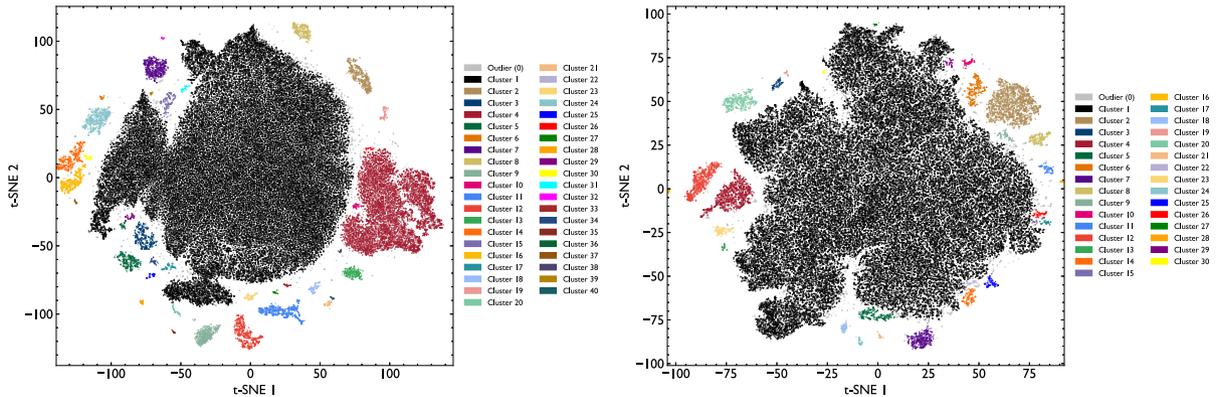

**Figure D3.** Embedding clusters for the *2D-PCA* case (left) and *2D-AE* case (right).

## APPENDIX E: TRANSIENT-DOMINANT CLUSTERS

Table E1 lists the transient-dominant clusters in the different embedding spaces used for the selection of transient candidates.

**Table E1.** Transient-dominant clusters used to find new transient candidates.

| Case | Flares | Dips |
|---|---|---|
| *2D-PCA* | 3, 6, 10, 15, 18, 19, 22, 23, 29, 31, 33, 36, 37 | 12, 35 |
| *3D-PCA* | 2–4, 6, 10, 11, 13–15, 17–19, 21, 22, 24 | – |
| *2D-AE* | 5, 6, 8, 10, 11, 14, 17, 21, 25, 26, 30 | 9 |
| *3D-AE* | 8 | 4 |

## APPENDIX F: CATALOGUE COLUMNS

Table F1 shows the column descriptions of the X-ray transient candidates catalogue.

**Table F1.** Column descriptions of the catalogue of X-ray transient candidates found in this work.

| Column name | Column description |
|---|---|
| CATALOG_ID | Catalogue identifier for the region event file obtained from the observation ID and region ID in the CSC |
| CSC_name | Name of the source in the CSC |
| TRANSIENT_TYPE | Transient type (F for flare candidate, D for dip candidate) |
| CSC_ra | Right ascension of the source in the CSC (°) |
| CSC_dec | Declination of the source in the CSC (°) |
| CSC_obs_id | Observation ID of the observation in the CSC |
| CSC_region_id | Region ID of the source in the CSC |
| CSC_significance | Highest flux significance in any energy band in the CSC |
| CSC_flux_aper_b | Background subtracted and aperture corrected broad-band flux in the CSC |
| CSC_hr_hm | Hard-to-medium energy band hardness ratio as defined in the CSC |
| CSC_hr_ms | Medium-to-soft energy band hardness ratio as defined in the CSC |
| CSC_hr_hs | Hard-to-soft energy band hardness ratio as defined in the CSC |
| CSC_var_prob_b | Variability probability in the broad energy band as defined in the CSC |
| CSC_var_prob_h | Variability probability in the hard energy band as defined in the CSC |
| CSC_var_prob_m | Variability probability in the medium energy band as defined in the CSC |
| CSC_var_prob_s | Variability probability in the soft energy band as defined in the CSC |
| CSC_var_index_b | Variability index in the broad energy band as defined in the CSC |
| CSC_gti_obs | Start date and time of the source observation in the CSC |
| CSC_gti_end | End date and time of the source observation in the CSC |
| CSC_theta | Off-axis angle of the source in the CSC (arcmin) |
| CSC_cnts_aper_b | Total counts measured in the modified source region for the broad energy band in the CSC |
| CSC_cnts_aperbkg_b | Total counts measured in the modified background region for the broad energy band in the CSC |
| CSC_type | Source type in the CSC (P for point sources, X for extended compact sources) |
| CSC2.1_flag | Flag for processing status in the CSC (True for CSC 2.1 processing completed, False for only CSC 2.0 processing completed) |
| PS_id | PanSTARRS-1 DR2 unique object identifier |
| PS_internal_id | PanSTARRS-1 DR2 internal object identifier |
| sep_CSC_PS | Angular distance between CSC and PanSTARRS-1 DR2 sources (arcsec) |
| PS_g | PanSTARRS-1 *g* magnitude (mag) |
| PS_r | PanSTARRS-1 *r* magnitude (mag) |
| PS_i | PanSTARRS-1 *i* magnitude (mag) |
| PS_z | PanSTARRS-1 *z* magnitude (mag) |
| GAIA_id | Gaia unique object identifier |
| sep_CSC_GAIA | Angular distance between CSC and *Gaia* DR3 sources (arcsec) |
| GAIA_g | Gaia *G* magnitude (mag) |
| GAIA_rv | Gaia radial velocity (km s$^{-1}$) |
| GAIA_parallax | Gaia parallax (mas) |
| GAIA_bp_rp | Difference in magnitude between the blue photometer (BP) and the red photometer (RP) measurements in *Gaia* (mag) |
| LS_id | Legacy Survey DR10 identifier |
| LS_object_id | Legacy Survey DR10 object identifier |
| sep_CSC_LS | Angular distance between CSC2.1 and Legacy Survey DR10 sources (arcsec) |

**Table F1** *– continued*

| Column name | Column description |
| --- | --- |
| LS_g | Legacy Survey DR10 $g$ magnitude (mag) |
| LS_r | Legacy Survey DR10 $r$ magnitude (mag) |
| LS_z | Legacy Survey DR10 $z$ magnitude (mag) |
| 2MASS_id | 2MASS name |
| sep_CSC_2MASS | Angular distance between CSC and 2MASS sources (arcsec) |
| 2MASS_j | 2MASS $J$ magnitude (mag) |
| 2MASS_h | 2MASS $H$ mag (mag) |
| 2MASS_k | 2MASS $K$ mag (mag) |
| SPEC_group_id | Group ID for SDSS DR17 spectra with multiple CSC matches |
| sep_CSC_SPEC | Angular distance between CSC and SDSS DR17 spectroscopic sources (arcsec) |
| SPEC_class | SDSS DR17 spectroscopic class (GALAXY, QSO, or STAR) |
| SPEC_subclass | SDSS DR17 spectroscopic subclass |
| SPEC_z | SDSS DR17 spectroscopy redshift |
| SPEC_zerr | SDSS DR17 spectroscopy redshift error |
| SPEC_wavemin | SDSS DR17 spectroscopy minimum observed (vacuum) wavelength (Angstroms) |
| SPEC_wavemax | SDSS DR17 spectroscopy maximum observed (vacuum) wavelength (Angstroms) |
| SPEC_sn | SDSS DR17 spectroscopy median signal-to-noise over all good pixels |
| SPEC_group_size | Number SDSS DR17spectra matched to CSC2.1 source in the group |
| SDSS_id | SDSS DR15 object identifier |
| sep_CSC_SDSS | Angular distance between CSC and SDSS DR15 sources (arcsec) |
| SDSS_type | SDSS DR15 type classification (S for stars, G for galaxies) |
| SIMBAD_ids | SIMBAD object identifier or catalogue names |
| sep_CSC_SIMBAD | Angular distance between CSC and SIMBAD sources (arcsec) |
| SIMBAD_otype | SIMBAD object type classification |
| SIMBAD_sptype | SIMBAD spectral type of the object |
| SIMBAD_parallax | SIMBAD parallax (mas) |
| SIMBAD_pmra | SIMBAD proper motion in right ascension (mas yr$^{-1}$) |
| SIMBAD_pmdec | SIMBAD proper motion in declination (mas yr$^{-1}$) |
| SIMBAD_rv | SIMBAD radial velocity (km s$^{-1}$) |
| SIMBAD_z | SIMBAD radial redshift |

This paper has been typeset from a T<small>E</small>X/L<small>A</small>T<small>E</small>X file prepared by the author.